\documentclass[12pt]{article}

\usepackage{graphicx}

\usepackage{amsfonts}

 \usepackage{url}

\begin{document}

\begin{center}

\Large{\bf Quasi Band-Limited Coronagraph for Extended Sources}

\vspace{1cm}

Igor Loutsenko and Oksana Yermolayeva

\vspace{5mm}

{\small Laboratoire de Physique Math\'ematique,\\
CRM, Universit\'e de Montr\'eal\\[1mm]}

\vspace{1cm}

\begin{abstract}

{

We propose a class of graded coronagraphic ``amplitude" image masks for a high throughput Lyot-type coronagraph that transmits light from an annular region around an extended source and suppresses light, with extremely high ratio, from elsewhere. The interior radius of the region is comparable with its exterior radius. The masks are designed using an idea inspired by approach due M.J. Kuchner and W.A. Traub (``band-limited" masks) and approach to optimal apodization by D.Slepian. One potential application of our masks is direct high-resolution imaging of exo-planets with the help of the Solar Gravitational Lens, where apparent radius of the ``Einstein ring" image of a planet is of the order of an arc-second and is comparable with the apparent radius of the sun and solar corona.

}

\end{abstract}

\end{center}

\begin{tabular}{ll}
\end{tabular}

\begin{figure}
\includegraphics[width=195mm]{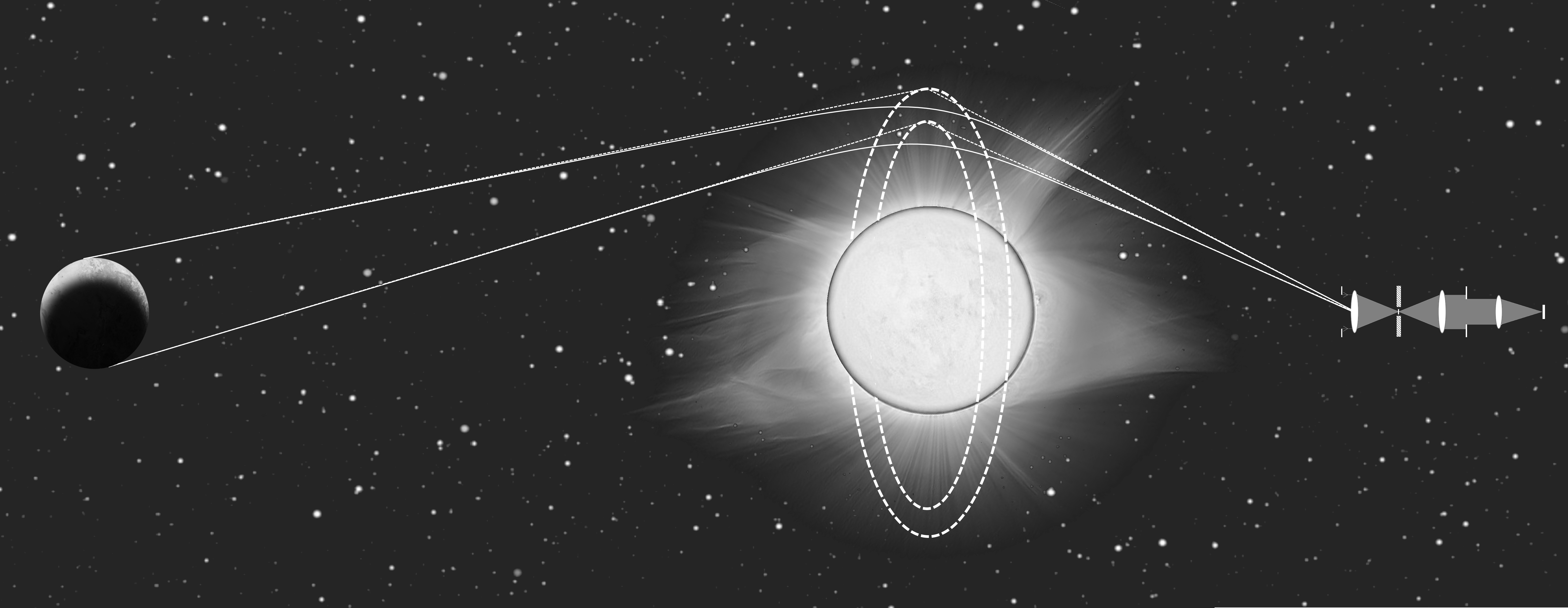}
\end{figure}

\vspace{1cm}

\section{ Introduction}

\vspace{5mm}

Recent discovery of thousands of exo-planets stimulates development in astronomical instrumentation and particularly in coronagraphy. The progress shown in coronagraphy gives us a hope for direct imaging of earth-like planets in the near future. However, high resolution imaging of those planets seems to be hardly possible due to their tiny apparent sizes: To observe such a planet at distance of about 100 ly with resolution of just few pixels one needs a telescope of aperture of an order of 100 km. Nature has presented us with a powerful ``instrument" that can resolve this problem. We mean the Solar Gravitational Lens (SGL), which focuses light from distant sources. The signal from the sun is far brighter than the gravitationally-lensed image of exo-planets, so some sort of coronagraph has to be used. Another difficulty is that the sun cannot be considered a point source, since the angular diameter of the solar disk is comparable to the separation between the disk and the planet's image.

The present article is mainly devoted to development of a new coronagraph instrument which resolves the above problem of suppression of light from ``small" \footnote{In examples considered in this article the apparent size of the source is between $1$ and $10$ diffraction limits.} extended sources. Although potential applications of our masks are not restricted only to SGL imaging, we will demonstrate main principles of the mask design in the context of SGL imaging, since the latter was our initial motivation. We note that a simple gaussian soft edge mask optimized for the SGL imaging has been already proposed in \cite{Tetal}. Whereas our mask provides extremely high rejection ratio as well as big throughput \footnote{Comparative analysis of performance of our mask with the gaussian soft edge mask is presented in the discussion section and Appendix 3.}, the principal purpose of the present paper is to demonstrate the power of the optimal band-limiting applied to coronagraphy, rather than to construct specific application to the SGL imaging.

The idea of using the sun as a powerful telescope goes back to Eshleman \cite{Eshleman}: The gravitational field of the sun acts as a spherical lens and magnifies intensity of electromagnetic radiation from distant objects along a semi-infinite focal line with the nearest point of observations being about $Z_{\rm min}$=550 AU (A good brief, self-contained introduction to the subject and related problems can be found in \cite{L}. For more details, one can e.g. see \cite{L1}, \cite{M}, \cite{TA}, \cite{TT} and references therein). For example, an integral intensity of radiation from an Earth-like exo-planet at distance 30 pc can be pre-magnified by the SGL up to six orders of magnitude (see e.g. \cite{L}). Theoretical angular resolution of the SGL is comparable to that of a telescope with aperture of the order of the sun size. At visible wavelengths this resolution could be as small as $10^{-10}$ arcsec (see e.g. \cite{TT}).

Recently, properties of the solar gravitational lens attracted attention both due to discovery of numerous exo-planets and the success of the Voyager-1 spacecraft, presently operating at about 140AU. Possibilities of high resolution (up to mega-pixel) imaging of such planets from the focal line of solar gravitational lens are now being discussed.

\begin{figure}
  \centering
  \includegraphics[width=175mm]{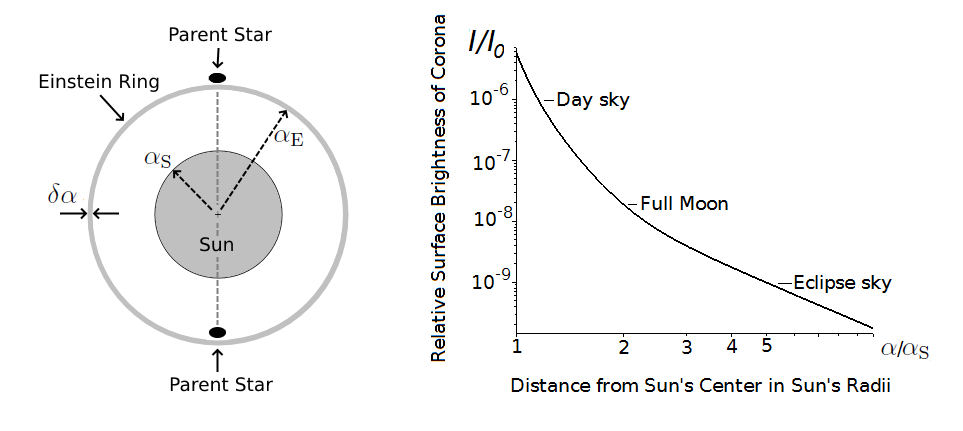}
  \caption{Left: Geometrical optics image of the sun, Einstein ring image of the planet and two images of the parent star. Right: Ratio of surface brightness of solar corona to surface brightness of the sun center.}
  \label{fig_SGL}
\end{figure}

Without going into much detail, we recall that an observer at distance $Z$ from the sun sees the image of an exo-planet as the ``Einstein ring" of the apparent radius $\alpha_{\rm E}$
$$
\alpha_{\rm E}(Z)=\alpha_{\rm max}\sqrt{Z_{\rm min}/Z},
$$
where $\alpha_{\rm max}=1.75$ arcsec, is the apparent radius of the sun at $Z=Z_{\rm min}=550$ AU. The width (i.e. apparent thickness) of the ring $\delta\alpha$ equals a half of the apparent (angular) diameter of an exo-planet (see Figure \ref{fig_SGL}).
For an Earth-like planet at 30pc from the sun $\delta\alpha\sim 10^{-6}$ arcsec. As a consequence, the Einstein ring width is not resolvable by any realistic telescope. In other words, for any practical purpose the ring can be considered as a circle whose brightness is varying along the circle circumference \footnote{The distribution of linear brightness along such a circle is essentially a Radon transform of the image of the planetary disc.}. A telescope with radius of aperture $\sim 1$m resolves the circumference of the ring with up to $\sim 10^{2}$ elements. Thus, performing a scan in the observer plane by taking a set of images of the Einstein rings from different $(X,Y)$ - positions, one could then make a tomographic reconstruction of the real image of a planet.
Note that amplification of the integral energy flow density by SGL is of order $2\alpha_{\rm E}/\delta\alpha$ (for details see e.g. \cite{L}).

Apart from formidable technical difficulties of getting to the focal line, there are principal problems of suppression of the diffraction glare from the sun and its corona. Indeed, the apparent radius of the sun at distance $Z$ is
$$
\alpha_{\rm S}(Z)=\alpha_{\rm max}Z_{\rm min}/Z.
$$
The separation between the solar disc and the Einstein ring $\alpha_{\rm E}(Z)-\alpha_{S}(Z)$ is zero at $Z=Z_{\rm min}$ and reaches its maximum at $Z=4Z_{\rm min }\approx 2200$ AU. The maximal separation equals $\alpha_{\rm max}/4\approx 0.44$ arcsec. At this distance the apparent radius of the sun $\alpha_{\rm S}$ equals the separation angle.

The distance $Z=4Z_{\rm min}\approx 2200$ AU is considered as minimal practical distance of the observation \cite{L}: The image of the ring will be superimposed with that of solar corona. The (angular) surface brightness of corona at the Einstein ring (i.e. at $\alpha_{\rm E}=\alpha_{\rm max}/2\approx 0.88$ arcsec) equals the surface brightness of the Earth night sky at full moon (astronomical
observations of faint objects are not taken during periods of full moon due to the sky
brightness). This surface brightness is about $10^{-8}I_0$, where $I_0$ is the surface brightness of the sun at its center (see Figure \ref{fig_SGL}).

More rigorous argument in favour of the above minimal practical distance is the following: The surface brightness of an Earth-like planet is about $10^{-5}$ times of that of its host star  \footnote{ The distance between the sun and the earth is about  $2\times 10^2$ sun's radii. By the inverse square law the light's flux density  at the earth orbit is about $4\times 10^4$ times smaller than that near the sun. The flux density scattered/reflected by the earth is about of the same order, i.e. about $10^{-5}$ of density at the sun surface".}. In geometrical optics the surface brightness is invariant. Therefore, the surface brightness of the geometrical optics image of the Einstein ring is $I_{\rm G}\sim 10^{-5}I_0$. However, the width of the ring is not resolved by a telescope, and one has to take into account the reduction of the surface brightness due to diffraction on the telescope aperture \footnote{For an optical system, the surface brightness is the energy flux per unit solid angle divided by the pupil area.} (i.e. ``diffractional widening" of the ring by several orders of magnitude) .

Take, for instance, a telescope comparable with the Hubble Space Telescope, i.e. a telescope with the diameter $D\approx 2.5$ m operating at visible wavelengths $\lambda<750$ nm. The characteristic (diffraction) width of the image of Einstein ring in the focal plane of the telescope is of order $\lambda/D$, which is about $10^{-1}$ arcsec. This is about $10^4-10^5$ times bigger than the geometrical optics width of the Einstein ring. Therefore, the surface brightness of the ring in the focal plane of the telescope $I$ will be $10^{-4}-10^{-5}$ times $I_{\rm G}$, i.e. about $10^{-9}-10^{-10}$ times $I_0$. Taking into account that the brightness of corona at $\alpha=\alpha_{\rm E}(4Z_{\rm min})$ is of order $10^{-8}I_0$, we see that it exceeds $I$ by one-two orders of magnitude. In other words, even at $Z=2200$ AU the image of the ring in the focal plane of a realistic space telescope will be one-two orders of magnitude fainter than that of the background corona \footnote{We note that the brightness of corona strongly increases towards the sun disc edge, from $\sim 10^{-8}I_0$ at two solar radii from the center of the sun to $\sim 10^{-5}I_0$ at one solar radius (see Figure \ref{fig_SGL}).}. That is why $Z=2200$ AU is considered as a minimal practical distance of observation.

From the above it follows that one may need to suppress the diffraction glare of the sun, so that the surface brightness of the glare (in vicinity of the ring image) in the final image plane of the optical system should be about $10^{-10}I_0$. It is important to stress that not only on-axis light, but also off-axis light with incidence angles $\alpha<\alpha_{\rm S}$ should be suppressed, while the light at $\alpha\approx\alpha_{E}$ should be transmitted almost entirely (we recall that $\alpha_{\rm S}=\alpha_{\rm E}/2$ at $Z=4Z_{\rm min}\approx 2200$ AU).

Below we propose a coronagraph that satisfies the above requirements. We will also
consider higher throughput designs with relaxed suppression conditions, namely, with the glare suppressed
to the level of corona $\sim 10^{-8}I_0$ (it is argued that this level of suppression might be sufficient for the SGL imaging \cite{Tetal}). In addition, our mask suppresses not only the light from the sun but also the light from most of corona, except the part of corona in a close vicinity of the ring.

In the next section we recall general principles of the Lyot-type coronagraphs as well as those with the ``band-limited" masks introduced by  M.J. Kuchner and W.A. Traub in \cite{KT}. Then we introduce a quasi band-limited amplitude mask \footnote{Under the ``amplitude mask" we mean a mask with real non-negative amplitude transmission factor, i.e. a mask which does not introduce a phase shift (see below).} using approach similar to that of optimal apodization by D. Slepian \cite{S}. Later, we consider the problem of suppressing light from the parent star of an exo-planet introducing ``one-dimensional" mask based on Slepian's solution of one-dimensional band-limiting problem. Finally, we introduce the ``product" mask which suppresses not only sunlight and the light from a part of corona, but also the light from the parent star. Tolerance to manufacturing errors is discussed in the concluding section of the paper. Analysis of suppression of light from solar corona can be found in Appendix 4.

\section{Band-Limited Mask}

\begin{figure}
  \centering
  \includegraphics[width=155mm]{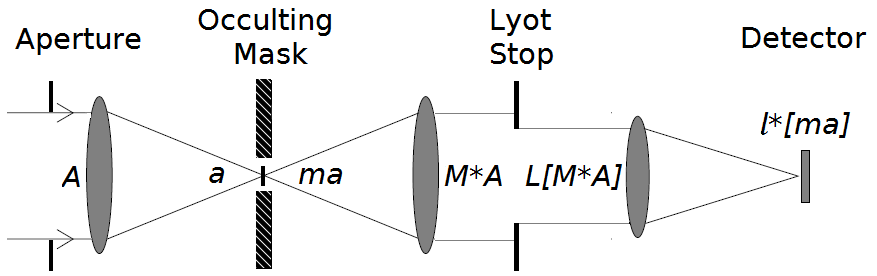}
  \caption{Schematic representation of principal stages of field propagation through coronagraph. Quantities denoted by the upper and lower case letters are related by the Fourier transform. For instance, $M$ is a Fourier transform of $m$ etc.}
  \label{stages}
\end{figure}

To establish notations, let us first briefly review general principles of the Lyot-type coronagraph (for more detail see e.g. \cite{SKMBK} and references therein). Stages of propagation of light through the coronagraph are depicted in Figure \ref{stages}.

We consider a telescope with primary of diameter $D$. Let $\vec{R}$ be two dimensional vector defining the coordinates in the entrance pupil plane. Then it is convenient to introduce dimensionless coordinates $\vec{x}=\vec{R}/D$. Similarly, let $\vec{r}$ be two-dimensional coordinates in the first image (focal) plane of the telescope. We re-scale them in the units of the diffraction characteristic scale $\lambda \mathcal{F}/D$, where $\cal{F}$ is the telescope focal length, introducing dimensionless coordinates $\vec{y}=\vec{r}D/(\lambda \cal{F})$. We will use the above coordinates as general coordinates of the coronagraph planes: $x$-coordinates for all pupil planes and $y$-coordinates for all focal planes. It is also convenient to re-scale the incidence angle $\vec{\alpha}$ of the plane wave in the units of the characteristic diffraction angle $\lambda/D$, introducing ``dimensionless" \footnote{In what follows we refer to $\vec{\alpha}$ as ``incidence angle" and to $\vec{\beta}$ as ``incidence".} incidence $\vec{\beta}=\vec{\alpha}D/\lambda$. Incidence $\vec{\beta}$ is a two dimensional vector whose cartesian coordinates we denote \footnote{Similar notations will be used for $x$ and $y$-coordinates, i.e. $\vec{x}=(x_\perp, x_\parallel)$ and $\vec{y}=(y_\perp, y_\parallel)$.} by $\beta_\perp$ and $\beta_\parallel$, i.e $\vec{\beta}=(\beta_\perp, \beta_\parallel)$.  These are dimensionless coordinates of a point source in the infinitely distant ``source" plane. The image of a point source in the first focal plane is centered at $\vec{y}=\vec{\beta}$ and here the $y$-coordinates of the image maximum coincide with the $\beta$-coordinates of the source.  One can think of $\beta$s as of the $y$-coordinates of the ``geometrical optics image" of the point source.

We consider a non-apodized entrance pupil, so, up to an $\vec{x}$-independent common factor, the dimensionless complex field of the plane wave of the unit amplitude immediately after encountering the primary mirror is
\begin{equation}
A_{\vec{\beta}}(\vec{x})=P(x)E_{\vec \beta}(\vec{x}),
\label{A}
\end{equation}
where
$$
E_{\vec \beta}(\vec{x})=e^{-2\pi i (\vec{\beta}\cdot\vec{x})}
$$
and $P$ is the non-apodized circular aperture function
$$
P(x)=\left\{\begin{array}{ll}
1, & x<1/2 \\
0, & x>1/2
\end{array} \right. , \quad x:=|\vec{x}|.
$$
Since components of $\vec{\beta}$ are not coordinates in the coronagraph planes we use a subscript-type function entry for this variable and do not put it in the common list of function entries \footnote{Later, we will move $\vec{\beta}$ in the common list of variables for the intensity related quantities such as point spread function (PSF), since integration of PSF wrt $\beta$ will be performed for non-point sources}.

From the Fraunhofer theory of diffraction it follows that amplitude of the field in the focal plane of the telescope is the Fourier transform of (\ref{A})
$$
a_{\vec{\beta}}(\vec{y})=\int A_{\vec{\beta}}(\vec{x})e^{2\pi i (\vec{x}\cdot\vec{y})}d^2x
$$
and
\begin{equation}
a_{\vec{\beta}}(\vec{y})=p\left(|\vec{y}-\vec{\beta}|\right), \quad p(y)=\frac{J_1(\pi y)}{2y},
\label{p}
\end{equation}
where $J_1$ stands for the Bessel function. In the Lyot-type coronagraph the image is focused on the occulting mask with amplitude transmission factor $m(\vec{y})$, so, immediately after passing the mask the field amplitude becomes
\begin{equation}
[ma](\vec{y})= m(\vec{y}) a_{\vec{\beta}}(\vec{y}) .
\label{am}
\end{equation}
The successive optics in the coronagraph transforms this product
to the second pupil plane, where the field is $M*A_{\vec{\beta}}$. Here, $*$ denotes convolution and $M(\vec{x})=\int m(\vec{y})e^{2\pi i (\vec{x}\cdot\vec{y})}d^2y$ is the Fourier transform of $m$. Note that we consider only symmetric graded masks, i.e. the masks with
$$
m(\vec{y})=m(-\vec{y}), \quad m=|m| .
$$
As a consequence, $M$ is real and also symmetric $M(\vec{x})=M(-\vec{x})$ in our case.

Now, the field passes through the Lyot stop, which can be described by an aperture function $L(x)$.
After passing the Lyot stop the field becomes
\begin{equation}
F_{\vec{\beta}}(x)=L(\vec{x})[M*A_{\vec{\beta}}](\vec{x}),
\label{L}
\end{equation}
where $[M*A_{\vec{\beta}}](\vec{x})=\int M(\vec{x'}-\vec{x})A_{\vec{\beta}}(\vec{x'})d^2x'$.

Then, the field passes the final pupil and is focused into the final (i.e. detector's) image plane where the field amplitude is the Fourier transform of (\ref{L}). In other words, the field amplitude in the final image plane equals
\begin{equation}
f_{\vec{\beta}}(\vec{y})=\left[l*[ma_{\vec{\beta}}]\right](\vec{y}),
\label{f}
\end{equation}
where $l$ is the Fourier transform of $L$. The field $F$ in (\ref{L}) is usually referred as the ``final field" and we refer its Fourier transform $f$ in (\ref{f}) as the ``detected field".  More precisely, detector registers intensity of image $\mu$ in the final (i.e. detector's) image plane. This intensity is the Point Spread Function (PSF) of the optical system and equals square of the absolute value of the detected field
$$
\mu(\vec{\beta},\vec{y})=|f_{\vec{\beta}}(\vec{y})|^2.
$$
The surface brightness $\mathcal{I}$ of a final image of a point source with incidence $\beta$ is the PSF divided by the Lyot stop area
\begin{equation}
\mathcal{I}(\vec{\beta},\vec{y})=\mu(\vec{\beta},\vec{y})/S, \quad S=\int L(\vec{x})d^2x .
\label{brightness_ps}
\end{equation}
For an extended incoherent source with surface brightness distribution $I_{\rm s}(\vec{\xi})$, the surface brightness at the detector plane $I$ equals\footnote{For a point source (i.e. plane wave) of a dimensionless unit intensity $I_{\rm s}(\vec{\xi})=\delta(\vec{\xi}-\vec{\beta})$.}
\begin{equation}
I(\vec{y})=\int I_{\rm s}(\vec{\xi})\mathcal{I}(\vec{\xi},\vec{y})d^2\xi .
\label{brightness}
\end{equation}
Now, we turn our attention to the band-limited masks. For such masks, the Fourier transform $M$ of the transmission amplitude $m$ has a finite support. In the rest of this and the next section we will consider only circularly symmetric masks $m(\vec{y})=m(y)$ (obviously, $M$ is also circularly symmetric). For such symmetric band limited mask, $M(x)$ is ``concentrated" completely within the disc of diameter $\epsilon$ and vanishes elsewhere. To distinguish a band-limited $M$ from a generic $M$ we introduce the special notation $M_\epsilon$ for the former, i.e.
\begin{equation}
M_\epsilon(\vec{x})=0, \quad x>\epsilon/2.
\label{bl}
\end{equation}
Its Fourier transform is denoted by $m_\epsilon$.

We also consider only circular non-apodized Lyot stops. Let dimensionless diameter of the stop be $\sigma<1$, then
$$
L(\vec{x})=P(x/\sigma), \quad S=\pi\sigma^2/4.
$$
It has been noticed in \cite{KT} that from (\ref{L}) and (\ref{bl}) it follows that in the case of Lyot stops, whose dimensionless diameter $\sigma$ does not exceed $1-\epsilon$, the final field equals
\begin{equation}
F_{\vec{\beta}}(\vec{x})=L(\vec{x})[M_\epsilon*A_{\vec{\beta}}](\vec{x})=L(\vec{x})[M_\epsilon*E_{\vec{\beta}}](\vec{x})=m_\epsilon(\vec{\beta})L(\vec{x})E_{\vec{\beta}}(\vec{x}) .
\label{Fb}
\end{equation}
In other words, the final field of the band-limited coronagraph \footnote{By the ``band limited coronagraph" we mean a Lyot-type coronagraph with the band limited mask and dimensionless diameter of the stop not exceeding $1-\epsilon$.} coincides with that of the plane wave of amplitude
$m_\epsilon(\vec{\beta})$ at incidence $\vec{\beta}$ that passed through a pupil of dimensionless diameter $\sigma\le 1-\epsilon$. Detailed derivation of (\ref{Fb}) is given in Appendix 1.

Thus, the detected field equals
$$
f_{\vec{\beta}}(y)=m_\epsilon(\beta)l(\vec{y}-\vec{\beta}) ,
$$
where for a circular stop of diameter $\sigma$
\begin{equation}
l(y)=\sigma^2 p(\sigma y)=\sigma\frac{J_1(\pi\sigma y)}{2y} .
\label{l}
\end{equation}

It follows that the intensity in the detector image plane equals $m_\epsilon(\beta)^2$ times the PSF of the stop \footnote{We recall that in our case $m$, $l$ and $f$ are all real.}, i.e
\begin{equation}
\mu(\vec{\beta},\vec{y})=m_\epsilon(\beta)^2 l^2\left(|\vec{y}-\vec{\beta}|\right)=m_\epsilon(\beta)^2\mu_{\rm Lyot}(\vec{\beta},\vec{y}),
\label{psf}
\end{equation}
where $\mu_{\rm Lyot}$ is the PSF of the Lyot stop.

Thus, the band-limited coronagraph completely suppresses an on-axis source when $m_\epsilon(0)=0$.

The final throughput of the energy for the incidence $\beta$ equals
\begin{equation}
\tau(\beta)=m_\epsilon(\beta)^2\sigma^2 .
\label{tau}
\end{equation}
Working throughput of the coronagraph equals $\tau(\beta_{\rm  W})$, where $\beta_{\rm W}$ is the working incidence. Usually one chooses $m_\epsilon\left(\beta_{\rm W}\right)$ to be close to unity in order to get the maximal useful throughput which is reached when $\sigma=1-\epsilon$.

It is important to note that the band limited coronagraph designed for the wavelength $\lambda$ will work for smaller wavelengths $\lambda'<\lambda$ as well, since the change $\lambda$ to  $\lambda'$ is equivalent to the scalings $m(y)\to m(\lambda' y/\lambda)$, $M(x)\to M(\lambda x/\lambda')$ and $\beta\to \lambda \beta/\lambda'$, while $\sigma$ remains invariant. As a consequence, the energy transmission at a given incidence angle $\alpha=\beta \lambda/D$ is the same for $\lambda'<\lambda$.

\section{Quasi Band-Limited Mask}

Our aim is to construct a coronagraph that transmits the light coming from an annular region containing the Einstein ring and suppresses the light coming from elsewhere. We recall that the band-limited coronagraph  completely suppresses light at incidences $\beta$ for which $m(\beta)=0$. However, the band-limited coronagraph that completely suppresses light coming from outside of the annular region is impossible to construct due to the uncertainty principle: A function and its Fourier transform cannot both have finite supports.

Below we propose construction of a graded ``amplitude" mask that is ``almost" band limited: Quasi Band Limited Mask or QBLM. Under ``amplitude mask" we mean that it is {\bf not} a phase changing on transmission mask. In other words, $m(y)$ is real and non-negative in our case. It is band-limited in the sense that the transmission factor
$m(y)$ has a finite support, vanishing outside some annual region of the interior radius $y_1$ and the exterior radius $y_2$
\begin{equation}
m(y)=0, \quad y\not\in y_1<y<y_2,
\label{annular}
\end{equation}
while its Fourier conjugate $M(x)$ is ``concentrated" in the disc of diameter $\epsilon$, ``almost" vanishing elsewhere. For instance, in examples that will be presented below, maximum of the ``tail" of $M$ outside the disc is about 5 orders of magnitude smaller than maximum of the ``main lobe" of $M$ on the disc (see Fig. \ref{fig_M}). In what follows, we refer to functions having the property (\ref{annular}) as ``annular-limited".

Les us now split $M(x)$ into two parts $M_\epsilon$ and $\delta M$:
\begin{equation}
M=M_\epsilon+\delta M  ,
\label{M_split}
\end{equation}
where $M_\epsilon$ is the ``main lobe" that vanishes outside the disc of diameter $\epsilon$
\begin{equation}
M_\epsilon(x)=0, \quad x>\epsilon/2,
\label{M_epsilon}
\end{equation}
and the ``tail", that vanishes on the disc
\begin{equation}
\delta M(x) =
0, \quad x<\epsilon/2
\label{delta_M}
\end{equation}
Taking a coronagraph with $\sigma\le 1-\epsilon$, from the decomposition (\ref{M_split}, \ref{M_epsilon}, \ref{delta_M}) and eqs. (\ref{L}, \ref{Fb}) we get the final field (for details see Appendix 1)
\begin{equation}
F_{\vec{\beta}}(\vec{x})=m(\vec{\beta})L(x)E_{\vec{\beta}}(\vec{x})-L(x)\Delta_{\vec{\beta}}(\vec{x}) ,
\label{FQ}
\end{equation}
where
$$
\Delta_{\vec{\beta}}(\vec{x})=\delta m(\vec{\beta})E_{\vec{\beta}}(\vec{x})-[\delta M*A_{\vec{\beta}}](\vec{x}) .
$$
Here, $\delta m$ is the Fourier transform of the ``tail" $\delta M$.

In the detector image plane we have
\begin{equation}
f_{\vec{\beta}}(\vec{y})=m(\vec{\beta})l\left(|\vec{y}-\vec{\beta}|\right)-\delta f_{\vec{\beta}}(\vec{y}) ,
\label{qbl}
\end{equation}
where
\begin{equation}
\delta f_{\vec{\beta}}=l*\left[(\delta_{\vec{\beta}}-a_{\vec{\beta}})\delta m\right],
\label{delta_f}
\end{equation}
and $\delta_{\vec{\beta}}$ stands for the two-dimensional Dirac $\delta$-function:
$$
\delta_{\vec{\beta}}(\vec{y})=\delta(\vec{y}-\vec{\beta}).
$$
We refer to the first term in the RHS of (\ref{qbl}) as the ``main field" , while the second term $\delta f$ is referred as the ``residue field".

Since the transmission amplitude vanishes outside the annulus (\ref{annular}), the main field for incidences outside this annular region
is completely suppressed and
$$
f_{\vec{\beta}}(y)=-\delta f_{\vec{\beta}}(y), \quad \mu(\vec{\beta},\vec{y})=\delta f_{\vec{\beta}}(y)^2, \quad \beta \not\in y_1<\beta<y_2 .
$$
The residue field is determined by the Fourier transform $\delta m$ of the ``tail". In what follows we call $\delta m$ the residue transmission. It is important to stress that residue transmission is an auxiliary mathematical value and not an amplitude transmission factor of the mask. Transmission factor of our mask is real and non-negative, i.e. our mask is not a phase mask, (i.e. it is pure ``amplitude mask"). From (\ref{M_split}, \ref{M_epsilon}, \ref{delta_M}) it follows that the residue transmission equals
\begin{equation}
\delta m(y) =m(y)-\hat{K}_\epsilon [m](y) ,
\label{mu}
\end{equation}
where $\hat{K}_\epsilon$ is the following integral operator
\begin{equation}
\hat{K}_\epsilon [m](\vec{y})=\int_{y_1<y'<y_2}K_\epsilon\left(\vec{y}-\vec{y'}\right)m(\vec{y'})d^2y', \quad K_\epsilon(\vec{y})=\epsilon\frac{J_1(\pi\epsilon y)}{2y}
\label{K_epsilon}
\end{equation}
and the integral is taken over the annulus $y_1<y<y_2$.

On the other hand, for working incidences, $m$ is of the order of unity and $m(\beta_{\rm W})\gg \max |\delta m(y)|$. As a consequence, at these incidences the residue field can be neglected and, similarly to the band limited coronagraph (\ref{psf}), PSF of the optical system is $m(\beta_{\rm W})^2$ times the PSF of the Lyot Stop.

\section{Optimal Occultation}

Now we are going to find the annular-limited $m(y)$, having the ``smallest" possible ``tail" $\delta M$. The tail is smallest in the sense of the optimal apodization by D.Slepian \cite{S}: that is, it has the smallest possible energy. In more detail, one has to maximize the energy of the main lobe, i.e. the ratio
\begin{equation}
\kappa=\frac{\int_{x<\epsilon/2} M(\vec{x})^2d^2x}{\int M(\vec{x})^2d^2x} ,
\label{kappa_x}
\end{equation}
where the integral in the numerator is taken over the disc of diameter $\epsilon$, while the integral in the denominator is taken over the whole plane. Obviously, $0<\kappa<1$. The above equation can be rewritten in terms of $m$:
\begin{equation}
\kappa=\frac{\int d^2y \int d^2y' K_\epsilon\left(\vec{y}-\vec{y'}\right) m(\vec{y})m(\vec{y'})}{\int m(y)^2d^2y'}, \quad
m(y)=0 \quad \textrm{if} \quad y\not\in y_1<y<y_2.
\label{kappa_y}
\end{equation}
It is not difficult to see that optimal $m(y)$ is an eigenfunction of the integral operator $\hat {K}_\epsilon$ corresponding to its maximal eigenvalue $\kappa$:
\begin{equation}
\kappa m(\vec{y})=\hat{K}_\epsilon [m](\vec{y}), \quad y_1<y<y_2 .
\label{km}
\end{equation}
We stress that $m(y)$ satisfies (\ref{km}) only on the interval $y_1<y<y_2$ (i.e. only within the annulus). It is not an eigenfunction on the whole plane. Since we consider radially symmetric $m(\vec{y})=m(y)$, the two-dimensional integral operator (\ref{K_epsilon}) can be reduced to one-dimensional one by integration in polar coordinates in the $y$-plane (see Appendix 2) and
\begin{equation}
\kappa m(y)=(\pi\epsilon)^2\int_{y_1}^{y_2}\mathcal{K}(\pi\epsilon y, \pi\epsilon y') m(y') y' dy', \quad y_1<y<y_2 ,
\label{K_reduced}
\end{equation}
where
\begin{equation}
\mathcal{K}(y,y')=\frac{yJ_1(y)J_0(y')-y'J_1(y')J_0(y)}{y^2-{y'}^2}.
\label{Kernel_1d}
\end{equation}
Our optimization differs from that of the apodization problem considered by D. Slepian \footnote{General formulation of the problem for arbitrary finite supports has been posed earlier, e.g. in \cite{S1964}. There general properties of $m$ are given, but a solution was presented only for the disc-limited case.} in \cite{S}. There, an analog of our function $m(y)$ is ``disc"-limited, which allows to reduce the apodization problem to solution of an eigenvalue problem for an ordinary differential operator. We do not know if a reduction to some ``sparce" operator is possible in the annular-limited case, but the diagonalization (\ref{K_reduced}) can be easily performed numerically \footnote{Diagonalization of our integral operator takes seconds of CPU time on modern PC, while at the time when works of D.Slepian et al \cite{S} were published such computation power was unavailable.} due to symmetry and positive definiteness of the operator.

\begin{figure}[t]
  \centering
  \includegraphics[width=175mm]{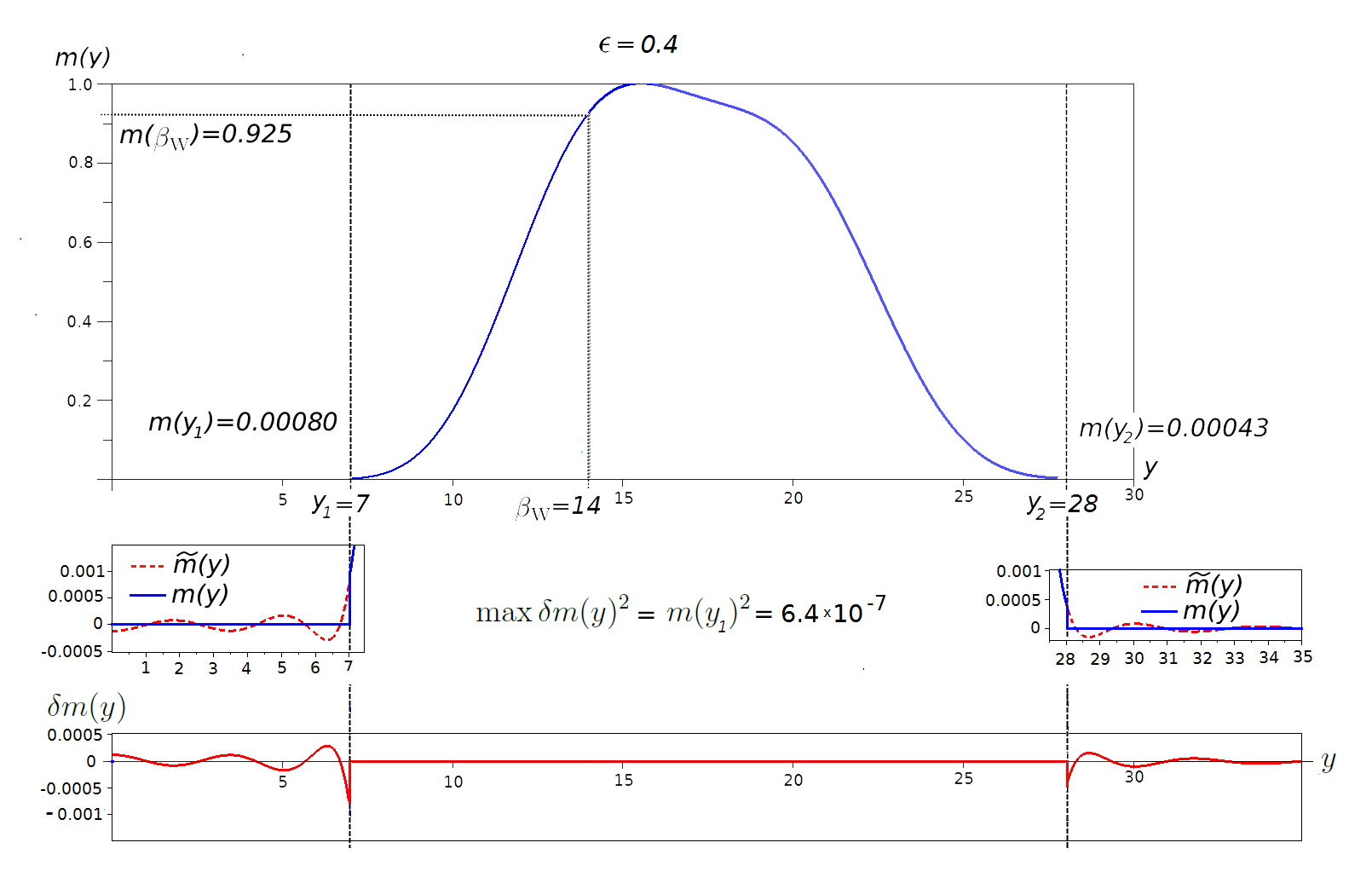}
  \caption{Top and middle rows: Optimal transmission amplitude $m(y)$ for the mask with $y_1=7$, $y_2=28$ and $\epsilon=0.4$ is represented by the solid line. Here $1-\kappa\approx 4\times 10^{-7}$. The working incidence $\beta_{\rm W}=2y_1=14$. The working throughput is 30 percents. We note that  formal continuation of transmission $\tilde m$ shown in the boxes in the middle row and $\delta m$ shown in the bottom row are auxiliary mathematical values and not the amplitude transmission factors of the mask. Transmission factor of our mask is real and non-negative, since our mask is not a phase mask (i.e. it is pure ``amplitude mask"). In difference from the continuation $\tilde m$ and residue transmission $\delta m$, the transmission factor $m$ vanishes outside the annulus $y_1<y<y_2$. Also, $m(y)$ is positive inside the annulus and discontinuous at the annulus boundaries, i.e. $m(y_1+0)>0$ and $m(y_2-0)>0$ (see graphs of $m(y)$ in the boxes of the middle row).}
  \label{fig_m}
\end{figure}

\begin{figure}[t]
  \centering
  \includegraphics[width=155mm]{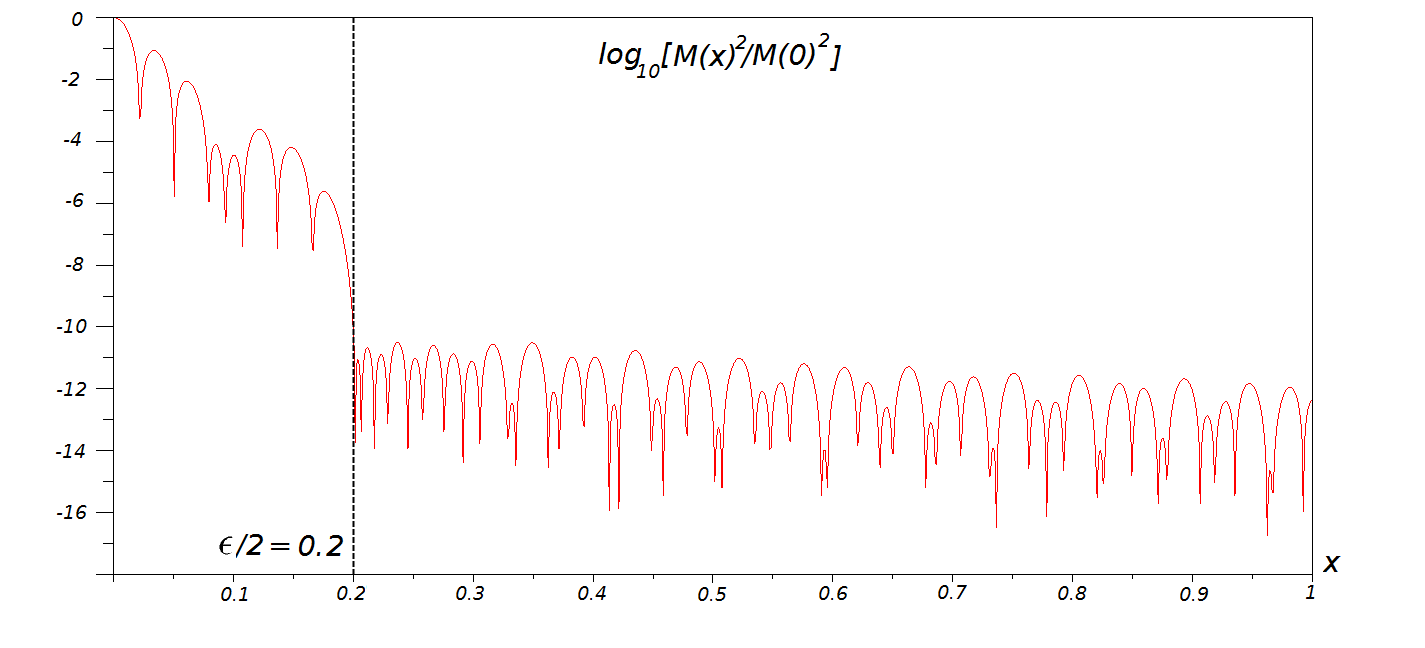}
  \caption{$\log_{10}\frac{M(x)^2}{M(0)^2}$ of the optimal mask from Figure \ref{fig_m} ($y_1=7$, $y_2=28$ and $\epsilon=0.4$). The part of the plot to the left from the vertical dashed line corresponds to the ``main lobe" of $M$, while the part on the right corresponds to its ``tail" (see eqs. (\ref{M_split}, \ref{M_epsilon}, \ref{delta_M})).}
  \label{fig_M}
\end{figure}

We call the problem of optimization of $m(y)$ for an arbitrary shaped support of the mask the optimal occultation. Similarly to the results of the optimal apodization, $\kappa$ is very close to $1$ in our examples of the optimal occultation (see below). There the value of $1-\kappa$ is of order of $10^{-7}$.

To make estimates of coronagraphic suppression, we will need to find the residue transmission $\delta m$. According to (\ref{annular}, \ref{mu}, \ref{km})
\begin{equation}
\delta m(y)=\left\{\begin{array}{ll}
(1-\kappa)m(y), & y_1<y<y_2 \\
-\hat{K}_\epsilon [m](y), & {\rm otherwise}
\end{array} \right .
\label{delta_m}
\end{equation}
It is worthy to note that $m(y)$ can be formally continued \cite{S1964} beyond the annular region by dropping the restriction $y_1<y<y_2$ in equation (\ref{K_reduced}). Then
$$
\tilde m(y)=\kappa^{-1}\hat{K}_\epsilon [m](y)
$$
equals $m(y)$ inside the annular region and continues it outside the region where, according to eq.(\ref{delta_m}), $\delta m=-\kappa \tilde{m}$. Since in all our examples $1-\kappa$ will be extremely small (of the order of $10^{-7}$), we can write that
$$
\delta m(y)\approx\left\{\begin{array}{ll}
0, & y_1<y<y_2 \\
-\tilde{m}(y), & {\rm otherwise}
\end{array} \right.
$$
and at boundaries of the region $\delta m(y_i)\approx -m(y_i), i=1,2$.

To get an idea of magnitude of the coronagraphic suppression, one can roughly estimate maximum of PSF for incidences $\beta$ outside the annular region and compare it with maximum of PSF with no mask applied (i.e. with maximum of PSF of the Lyot stop).

From (\ref{delta_f}) it follows that absolute value of the residue field $|\delta f_{\vec{\beta}}|$ is not exceeding the biggest of the two following values
\begin{equation}
2|\delta m(\beta)|\max |l|, \quad 2\max \left| l*\left[a_{\vec{\beta}}\delta m\right]\right| .
\label{max_residue}
\end{equation}
One may assume that these two values are of the same order on average and
$$
|\delta f_{\vec{\beta}}|<\mathcal{O}(\max |\delta m(y)|)\max |l| .
$$
Therefore, for incidences outside the annular region:
\begin{equation}
\max \mu(\vec{\beta},\vec{y})= s(\beta) \max \mu_{\rm Lyot}=s(\beta)\left(\frac{\pi \sigma^2}{4}\right)^2, \quad s(\beta)<\mathcal{O}(\max \delta m(y)^2), \quad \beta \not\in y_1<\beta<y_2,
\label{sb}
\end{equation}
where $\mu_{\rm Lyot}$ stands for PSF of the Lyot stop. In other words, according to our estimate the suppression coefficient $s(\beta)=\max \mu(\vec{\beta},\vec{y})/\max \mu_{\rm Lyot}$ is of order $\max \delta m^2$ or smaller order. This estimate is confirmed by direct numerical simulations of coronagraph which are presented in the next section. In fact, the actual suppression coefficient turns to be about two-three orders of magnitude smaller (i.e. suppression is 2-3 orders stronger)  than our very rough estimate $\max \delta m^2$  for incidences that are not in vicinity of boundary of the annular region (see Figures \ref{fig_points}, \ref{fig_suppression}). In vicinity of the boundary it is of the same order. A more accurate estimate could use some envelope $\tilde m_{\rm e}(\beta)$ of the oscillating function $\tilde m(\beta)$. The suppression coefficient and square of this envelope are of the same order  $s(\beta)=\mathcal{O}\left(\tilde m^2_{\rm e}(\beta)\right)$ (see Figure \ref{fig_suppression}).

Now, to be specific, we consider an example from the introduction section: Take $D=2.5$m telescope at $Z=4Z_{\rm min}$ and $\lambda=750$ nm. For these parameters the magnitude of the incidence corresponding to the Einstein ring (working incidence) $\beta_{\rm W}=\alpha_{\rm E}D/\lambda=14$. Let the apparent boundary of the sun coincide with the interior boundary of the annular region. Then $y_1=\beta_{\rm W}/2=7$. We chose the outer boundary of the region to be $y=y_2=2\beta_{\rm W}=28$.

We choose $\epsilon=0.4$ which corresponds to the maximal throughput $(1-\epsilon)^2=0.36$. The optimal transmission amplitude obtained by numerical solution of (\ref{K_reduced}) is shown on Figure \ref{fig_m}. The difference between the ``main lobe" and the ``tail" of the Fourier transform of $m$ is demonstrated on Figure \ref{fig_M}. The formal continuation $\tilde m(y)$ of $m$, is shown in separate boxes of Figure \ref{fig_m}. For this solution $1-\kappa\approx 4\times 10^{-7}$. Therefore, outside the annular region, $\delta m\approx-\tilde m$ with the relative precision $4\times 10^{-7}$. The residue transmission reaches maximum by absolute value at $y=y_1$ and $\max \delta m^2\approx 6.4\times 10^{-7}$. Therefore, one can expect the suppression to be of six or more orders of magnitude.

The transmission amplitude at working angle $m(\beta_{\rm W})$ equals $0.925$, so, according to (\ref{tau}) the maximal working throughput is $m(\beta_{\rm W})^2(1-\epsilon)^2\approx 0.308$ (i.e. about 30 percents).

\section{Results of Direct Numerical Simulations}

\begin{figure}
  \centering
  \includegraphics[width=175mm]{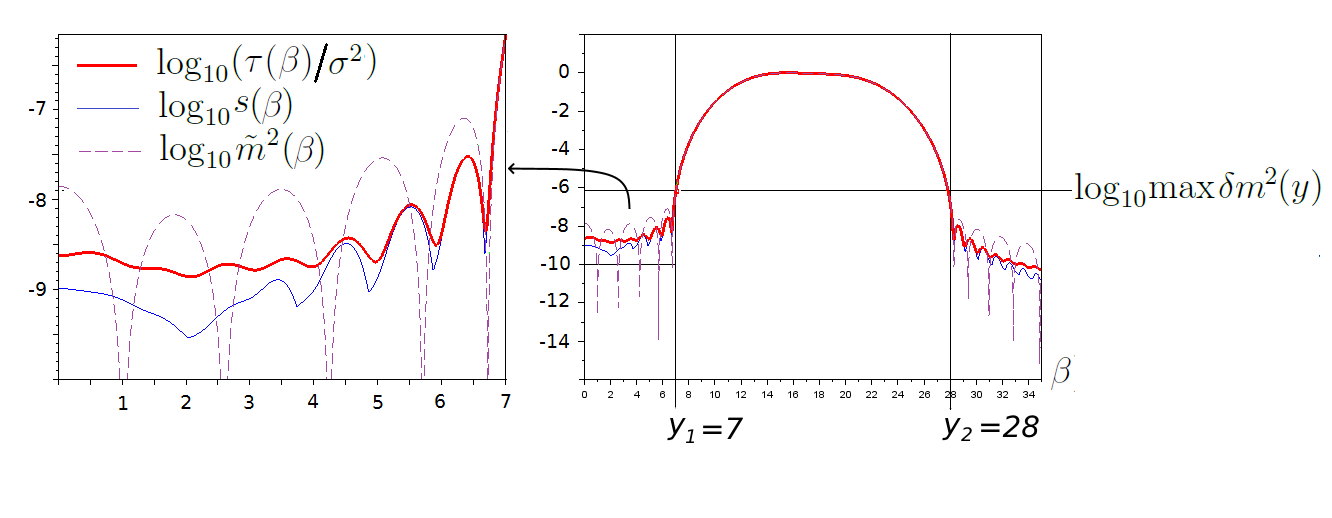}
  \caption{Thick solid curve: $\log_{10}$ of energy suppression coefficient $\tau(\beta)/\sigma^2$ (the latter is the ratio of energy passed with mask to that passed without mask). Thin solid curve: $\log_{10}$ of suppression of maximum of PSF at given $\beta$ (i.e. $\log_{10} s(\beta)$, see eq. (\ref{sb})). Dashed curve: $\log_{10}$ of square of continuation $\tilde{m}(\beta)$ (recall that $\tilde{m}(\beta)^2\approx\delta m(\beta)^2$ for $\beta \not\in y_1<\beta<y_2$).}
  \label{fig_suppression}
\end{figure}

Direct numerical simulations of a Quasi Band Limited Coronagraph are in agreement with the above estimates. We performed direct simulations of all stages of the field propagation (see eqs. (\ref{A}-\ref{f}) and Figure \ref{stages}) applying the fast Fourier transform at each pupil. Maximal size of the grid was $4096 \times 4096$, but sizes $2048 \times 2048$ and $1024 \times 1024$ give results which differ only by few percents from those of maximal grid. Simulations are run for the Lyot stop of the diameter $\sigma=1-\epsilon=0.6$.
\begin{figure}[t]
  \centering
  \includegraphics[width=175mm]{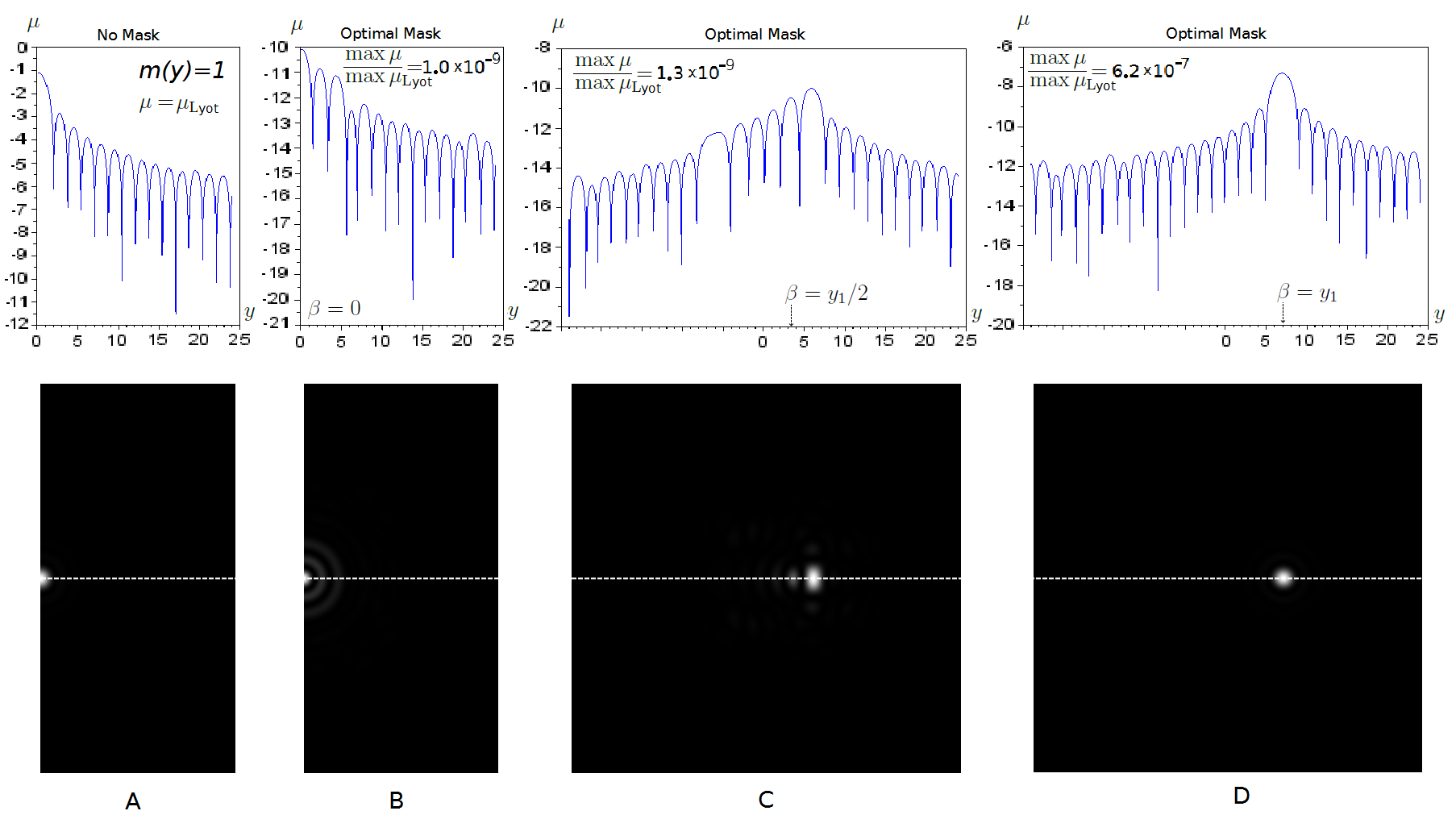}
  \caption{Bottom: Normalized images of point sources.  Top: $\log_{10}$ of PSF in the detector plane (i.e. $\log_{10}\mu$) along the central sections (i.e. along the dashed lines at corresponding bottom images). Panel A: No mask is applied. Panels B to D: optimal mask is applied. Corresponding incidences are: $\beta=0$ (Panel B, on-axis source), $\beta=y_1/2=3.5$ (Panel C) and $\beta=y_1=7$ (Panel D, source at interior boundary of annular region).}
  \label{fig_points}
\end{figure}
The suppression is minimal (i.e. suppression coefficients are maximal) in the neighborhood of interior boundary of the annular region. With a good precision the suppression coefficients equal $\delta m(\beta)^2$ in this neighborhood. The above value has its maximum $6.4\times 10^{-7}$ at the boundary and falls rapidly by about two orders of magnitude as $\beta$ decreases: The suppression coefficients are of the order of $10^{-8}-10^{-9}$ at the bigger part of the disc $\beta<y_1$, see Figure \ref{fig_suppression}. The suppression coefficient of maximum of PSF is of the same order as the coefficient of energy suppression $\tau(\beta)/\sigma^2$. The latter coefficient equals the ratio of energy flow passed in the presence of mask to that without mask.

The detector plane images of point sources at different incidences $\beta\le y_1$ and sections of intensity are shown on Figure \ref{fig_points}.

To evaluate the sun glare one has to compute integral (\ref{brightness}) of the product of distribution of the surface brightness of the sun with the PSF in the detector plane divided by area of the Lyot stop. We recall that we choose the parameters in such a way that apparent radius of the sun equals $y_1$ in units of characteristic diffraction angles, i.e. edge of the sun disc coincides with interior boundary of the annulus. In our computations we presented the sun as a disc of uniform surface brightness $I_0$, so that the limb darkening is not taken into account. Accounting for limb darkening will give better results \footnote{The surface brightness of the edge of the sun disc is less than half of that of the disc center. Since the boundary regions contribute most to the glare, the surface brightness of the glare obtained with account of the limb darkening will be approximately half of that we obtained for the uniform disc.} (i.e. the glare brightness will be smaller).
\begin{figure}
  \centering
  \includegraphics[width=175mm]{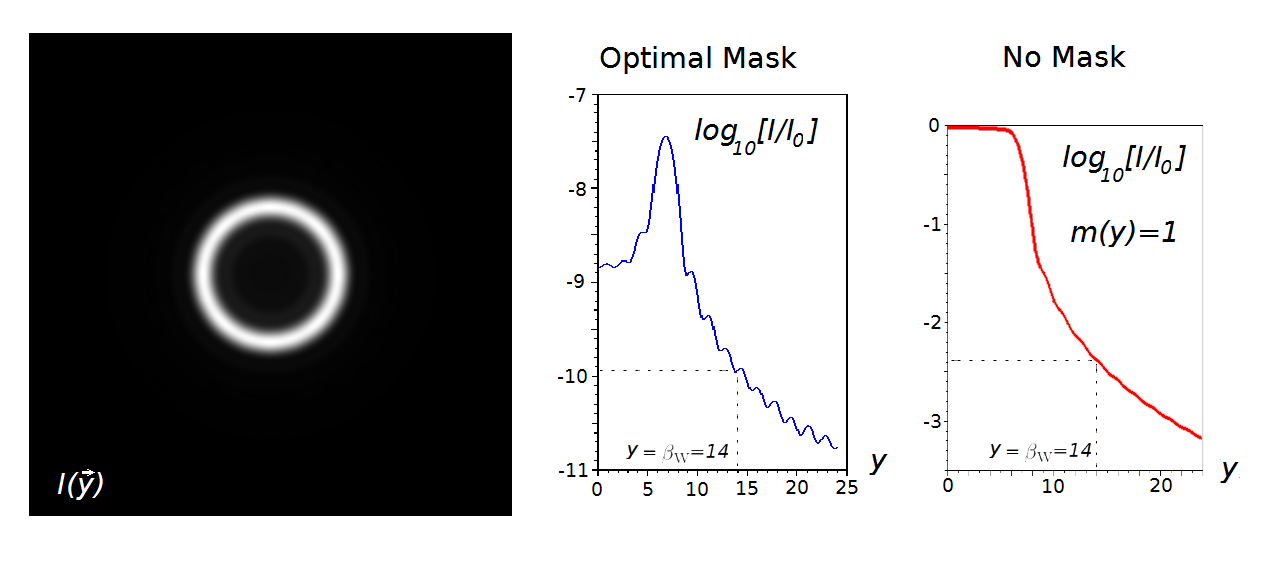}
  \caption{Left: Detector plane image of the solar disc. Optimal mask is applied. No limb darkening is taken into account (the sun is presented as a disc of uniform surface brightness $I_0$). Center: Radial dependence of the relative surface brightness in the detector plane. Right: Radial dependence without mask applied.}
  \label{fig_disc}
\end{figure}
With the optimal mask applied, the glare brightness at working angle $\beta_{\rm W}=2y_1=14$ (i.e. at position of the Einstein ring) is about $10^{-10}I_0$ (see Figure. \ref{fig_disc}). Therefore, the goal set in the introductory section can be achieved with help of our mask.

\section{Suppressing the Parent Star, One-dimensional Slepian's Mask, Product Mask}

The circularly symmetric mask can suppress the diffraction glare of the sun to the level of brightness of the detector plane image of an Earth-like exo-planet at about 30pc. However, there remains a problem of suppressing glare from the planet's parent star. The SGL produces two images of the parent star on the line that passes through the center of the sun (see Figure \ref{fig_SGL}). They appear at opposite sides wrt the center, one inside the circle $\alpha=\alpha_{\rm E}$ (i.e. inside circle $\beta=\beta_{\rm W}$) and the other outside it. Minimal angular distance between images and the circle is about a half of the apparent separation between the planet and its parent star. The size of the star is not resolvable by the telescope, so, for any practical purpose the star can be considered as a point source.

Since the SGL amplifies intensity of radiation from a planet much more stronger than it does from its parent star and apparent diameter of the Einstein ring is relatively big, an acceptable suppression level for observations with SGL is several orders of magnitude weaker than that required for observations without it. Indeed, the ratio of the amplification of radiation from an exo-planet to that from its parent star is of the order of the ratio of planet's orbit radius to its own radius. This is about $3\times 10^4$. Taking into account that the circumference of the Einstein ring in the detector plane is about $30$ resolution elements, reduction of the flux ratio \footnote{We define flux ratio as the ratio between the energy flow from the parent star and the energy flow per resolution element from an exo-planet.}
by the SGL is $\sim 10^3$. Since, for an Earth-like planet the unreduced flux ratio is about $\sim 10^{10}$, it will be $\sim 10^7$ in presence of the SGL. Also, since the apparent distance between the bigger part of the Einstein ring and images of the parent star is of the order of $10$ diffraction angles, acceptable magnitude of the suppression coefficients can be several orders bigger than $10^{-7}$ (we recall that, according to our definitions, smaller suppression coefficients mean bigger suppression and vice versa).

\begin{figure}
  \centering
  \includegraphics[width=135mm]{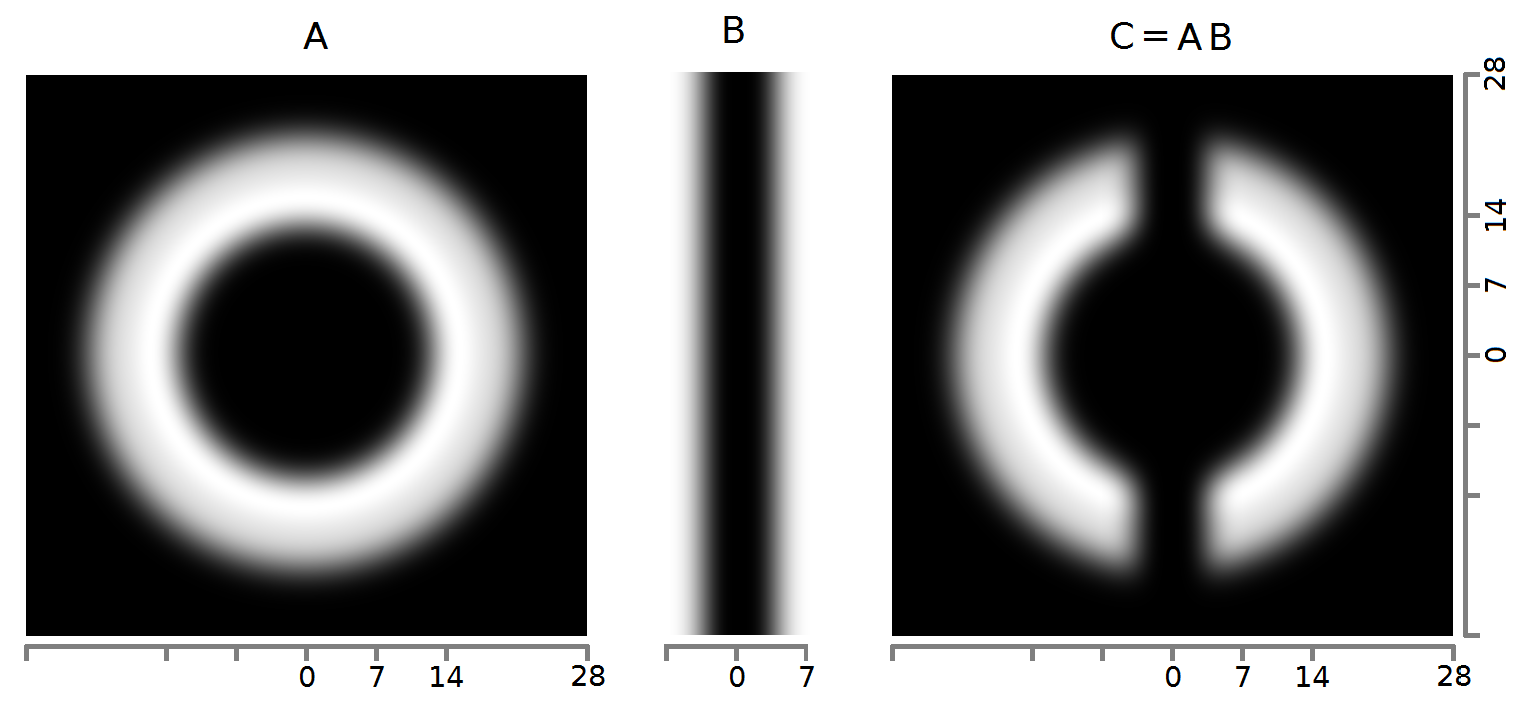}
  \caption{Left: Circularly symmetric mask. Center: One-dimensional ``Slepian's" mask. Right: Product mask. Gray-scale values of image pixels are proportional to the mask transparency $t=m^2$ (also called intensity transmission factor). Black color corresponds to $t=0$, while white color corresponds to $t=1$. The interior/exterior radii of the annular support are $y_1=7$ and $y_2=28$ correspondingly. The half-width of the Slepian's mask equals $y_1$. For both masks $\epsilon=0.4$. Einstein ring radius $\beta_{\rm W}=2y_1=14$.}
  \label{fig_masks}
\end{figure}

To solve the problem of simultaneous suppression of the sun and the parent star glare, we first introduce an effectively one-dimensional mask that eliminates light from sources located on the line passing through two images of the parent star. The transmission amplitude of this mask is essentially Slepian's solution to the one-dimensional optimal apodization problem \cite{S}. We call the corresponding mask Slepian's mask. Then we will test the mask that is the product of circularly symmetric mask (introduced in previous sections) and the Slepian's mask (see Figure \ref{fig_masks}).

Let $y_\perp$ and $y_\parallel$ be coordinates in the $y$-plane, i.e. $\vec{y}=(y_\perp,y_\parallel)$. We consider a mask with the transmission amplitude changing only in one direction $m=m(y_\perp)$. Our aim is to suppress point sources at the line $y_\perp=0$, so that $m(0)=0$. Also
$$
m(\vec{y})=1-u(y_\perp) ,
$$
where $u$ is an even function $u(y_\perp)=u(-y_\perp)$, such that $u(0)=1$ and $u(y_\perp)$ has a finite support $y_1<y_\perp<-y_1$:
$$
u(y_\perp)=0, \quad |y_\perp|>y_1.
$$
In other words, the Slepian's mask is completely transparent outside the strip $|y_\perp|<y_1$ ($m=1$ for $|y_\perp|>y_1$). The Fourier transform $M$ of $m$ is
$$
M(\vec{x})=\left[\delta(x_\perp)-U(x_\perp)\right]\delta(x_\parallel) ,
$$
where $U(x_\perp)=\int e^{2\pi i x_\perp y_\perp}u(y_\perp)dy_\perp$ is one-dimensional Fourier transform of $u(y_\perp)$ and $\vec{x}=(x_\perp,x_\parallel)$.
\begin{figure}[t]
  \centering
  \includegraphics[width=175mm]{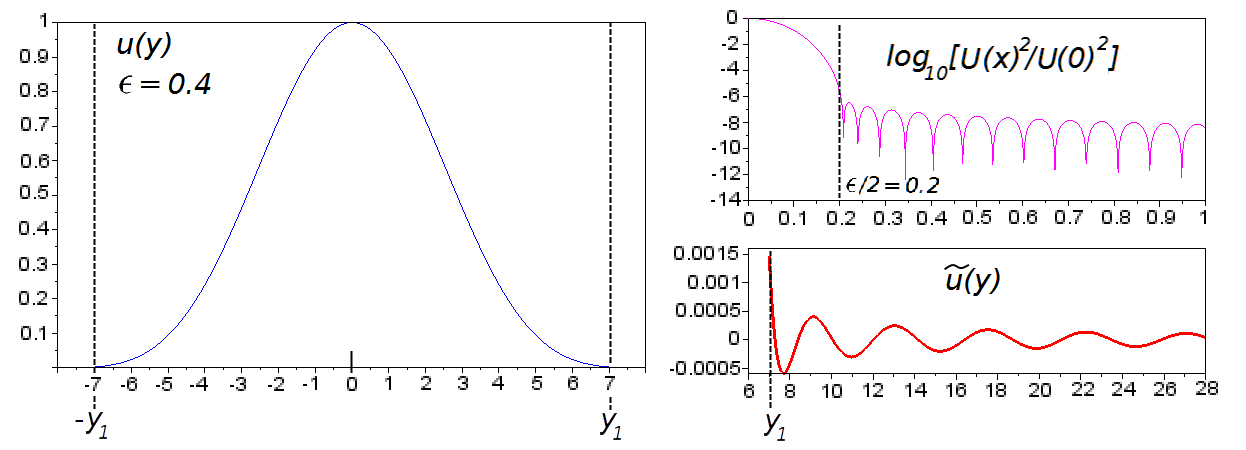}
  \caption{Left: Optimal $u(y)$ for the Slepian's mask with $y_1=7$ and $\epsilon=0.4$. Here $1-\kappa\approx 5\times 10^{-7}$. Top right: $\log_{10}\frac{U(x)^2}{U(0)^2}$. Bottom right: Formal continuation $\tilde{u}(y)$ of $u(y)$ (We recall that $\tilde{u}(y)\approx\delta m(y)$ when $|y|>y_1$.)}
  \label{fig_S}
\end{figure}
Similarly to the two-dimensional optimal occultation problem, we split $U(x)$ into the ``main lobe" $U_\epsilon(x)$ with support on the interval $-\epsilon/2<x<\epsilon/2$ and the ``tail" $\delta U(x)$:
$$
U=U_\epsilon+\delta U, \quad U_\epsilon(x)=\left\{\begin{array}{ll}
U(x), & |x|<\epsilon/2 \\
0, & |x|>\epsilon/2
\end{array}
\right., \quad
\delta U(x)=\left\{\begin{array}{ll}
0, & |x|<\epsilon/2 \\
U(x), & |x|>\epsilon/2
\end{array}
\right.
$$
Energy of the ``tail" is minimal when $u$ is an eigenfunction corresponding to the highest eigenvalue $\kappa$ of the one-dimensional integral operator:
$$
\kappa u(y)=\hat{K}_\epsilon[u](y)=\int_{-y_1}^{y_1}\frac{\sin\pi\epsilon (y-y')}{\pi(y-y')}u(y')dy' .
$$
The solution of the above equation for $\epsilon=0.4$ and $y_1=7$ (same values as in the circularly symmetric case) is shown on Figure \ref{fig_S}. Here the deviation of $\kappa$ from unity equals $5\times 10^{-7}$ and is of the same order as that of the example considered in previous section for the circularly symmetric case. The maximal residue of the transmission amplitude $\max|\delta m|\approx u(y_1)$ and is approximately of the same order as in the circularly symmetric example ($u(y_1)\approx 1.5\times 10^{-3}$, compare Figures \ref{fig_m} and \ref{fig_S}). Therefore, one can expect the same level of suppression for a point source at the line $\beta_\perp=0$. And, indeed, direct numerical simulations of a coronagraph with the Slepian's mask and a circular stop of the diameter $1-\epsilon=0.6$ give the result $\frac{\max \mu}{\max \mu_{\rm Lyot}}\approx  5.25 \times 10^{-10}$ for PSF and the energy suppression coefficient is equal $\approx 2\times 10^{-9}$ when the point source is on the line $\beta_\perp=0$. Obviously, the suppression coefficients do not depend on $\beta_\parallel$ for this one-dimensional mask.
\begin{figure}
  \centering
  \includegraphics[width=175mm]{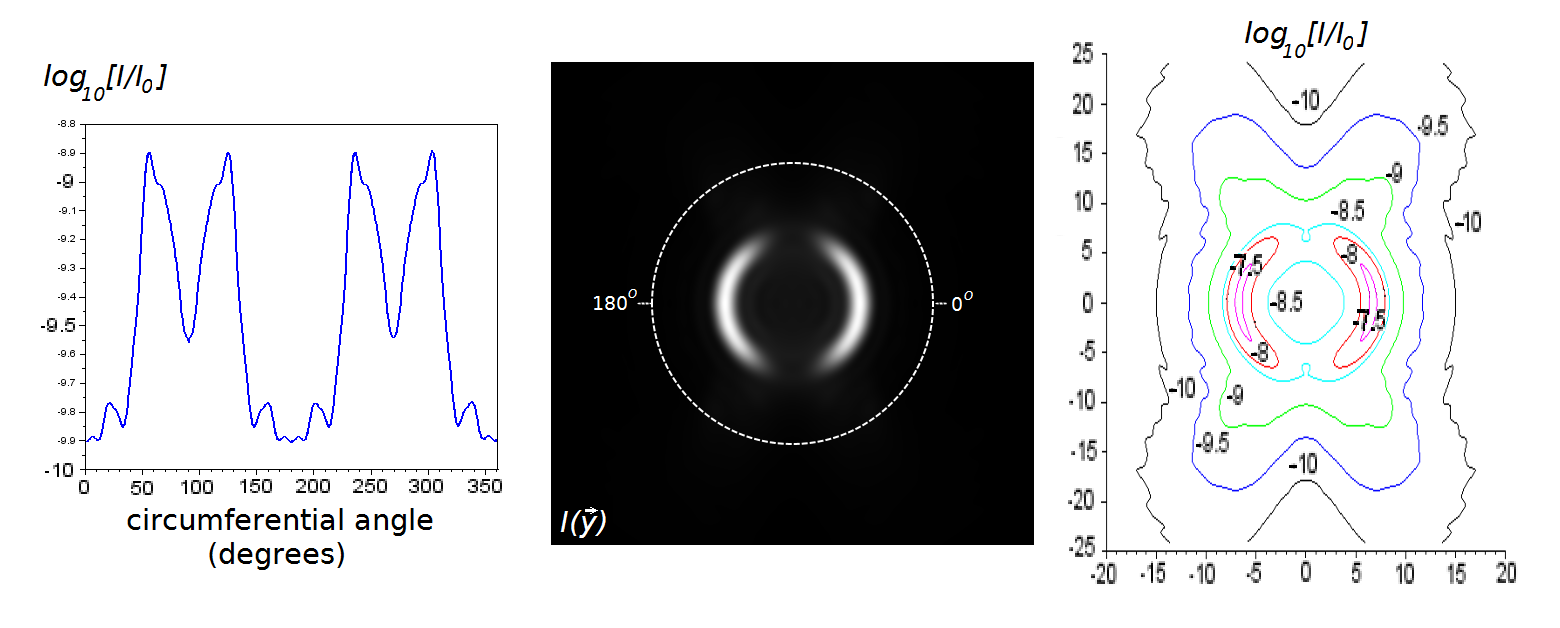}
  \caption{Center: Detector plane image of the solar disc. Product mask is applied. No limb darkening is taken into account (the sun is presented as a disc of uniform surface brightness $I_0$). Left: $\log_{10}$ of relative surface brightness $I/I_0$ along the circumference of the image of the Einstein ring $y=2y_1=14$. Right: Level contours of $\log_{10}$ of the relative surface brightness.}
  \label{fig_product_disc}
\end{figure}

Consider now the product mask, i.e. the mask whose amplitude transmission factor is the product of factors of the circularly symmetric and Slepian's mask (see Figure \ref{fig_masks}). The idea of using the product mask comes from the pure band limited case: A product of two band-limited functions $f(\vec{y})$ and $g(\vec{y})$ is also a band limited function, since the support of the convolution $[F*G](\vec{x})$ is a cartesian sum of supports of $F$ and $G$. In the case of product of the circularly symmetric and one dimensional masks, this is the cartesian sum of a (two-dimensional) disc and a (one-dimensional) segment. However, in the case of the quasi band limited masks, the support of new ``main lobe" can be smaller than the cartesian sum of supports of the ``main lobes" of $F$ and $G$. Indeed, the definition of support of the ``main lobe" as a region where most of the energy is concentrated is somehow arbitrary. Also, the convolution of ``main lobes" can be of the same order as the ``tail" on a substantial part of a cartesian product. That is why one can try to apply the product mask without changing the size of the Lyot stop.

We have performed direct numerical simulations of the coronagraph with the product mask without changing diameter of the stop $\sigma=1-\epsilon=0.6$, preserving $30$ percent working throughput. The results of simulations do not show substantial degradation in the suppression coefficients: The intensity of glare over the bigger part of the working region is of the same order as in the circularly symmetric case. The results of suppression of the glare from the sun are shown on Figure \ref{fig_product_disc}: Although, in difference from the circularly symmetric case, the working space is reduced, the glare at the bigger part of the Einstein ring (in total, more than $180$ degrees of the working space along the ring circumference) is of the same order as in the circularly symmetric case.
\begin{figure}
  \centering
  \includegraphics[width=175mm]{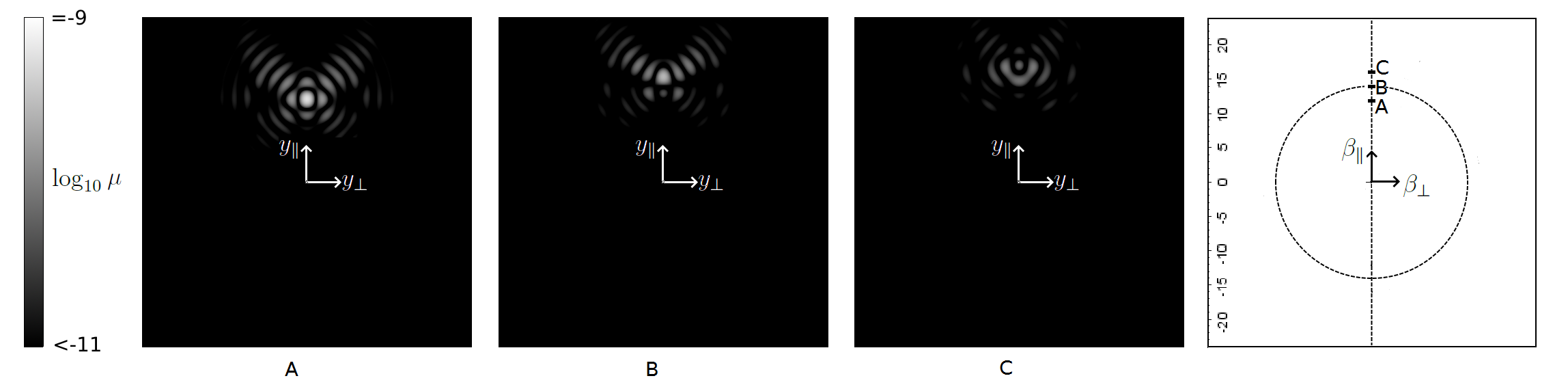}
  \caption{$\log_{10}$ of the detector plane PSF for the product mask and different position of point sources along the line $\beta_\perp=0$. Corresponding positions of the sources and the Einstein ring are shown at the rightmost panel ($\beta_\parallel=12$, $\beta_\parallel=14$ and $\beta_\parallel=16$ for the point sources at A, B and C correspondingly).}
  \label{fig_parent}
\end{figure}
\begin{figure}
  \centering
  \includegraphics[width=175mm]{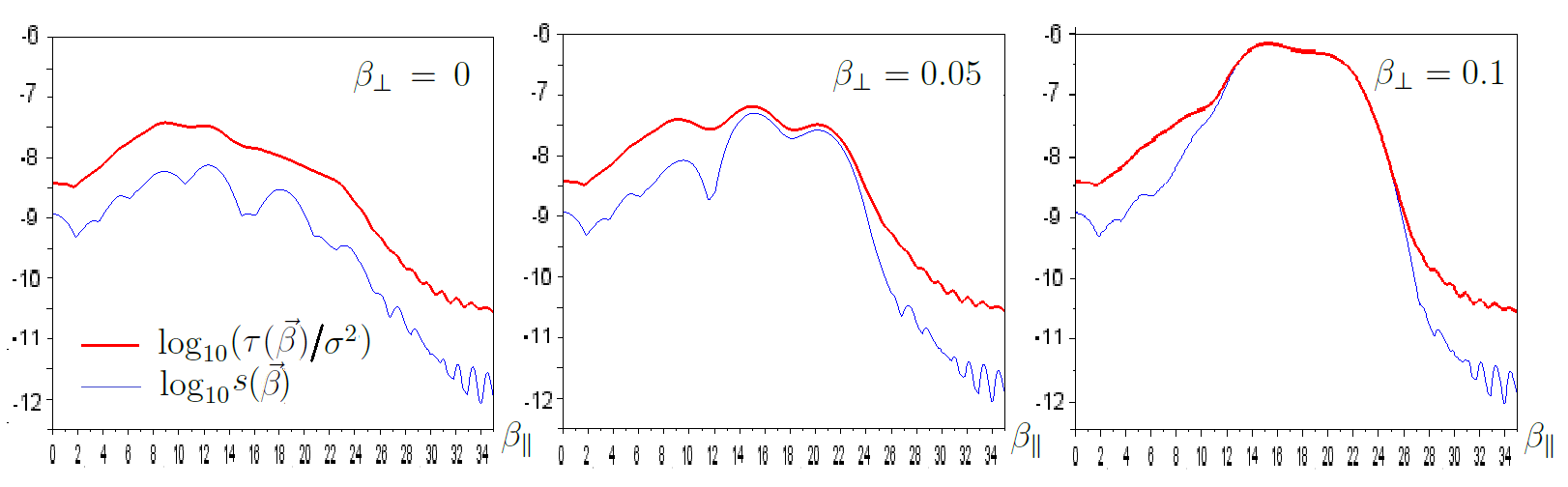}
  \caption{$\log_{10}$ of suppression coefficients of the product mask as functions of $\beta_\parallel$ for $\beta_\perp=0$, $\beta_\perp=0.05$ and $\beta_\perp=0.1$ (from left to right correspondingly). Thick line corresponds to the energy suppression coefficient, while thin line stands for the PSF maximum suppression coefficient.}
  \label{fig_suppression_product}
\end{figure}

Finally, the suppression coefficient for the point sources located at the line $\beta_\perp=0$ also do not show a substantial degradation: The level of suppression is more than sufficient for reducing glare from the parent star to an acceptable level. We also performed simulations of cases with off-alignment of the parent stars and the mask, i.e. the cases when $\beta_\perp\not=0$ (see Figure \ref{fig_suppression_product}). The acceptable deviation from the central line is at least of the order of $0.1$, (i.e. $\sim 1/10$ of the characteristic diffraction angle).

\section{Discussion and Conclusions}
\vspace{5mm}

In the present article we proposed graded coronagraph masks that can substantially suppress the diffraction glare from extended sources. In examples considered, the diffraction glare from a source in the form of disc of the apparent diameter of the order of one arc-second and of the uniform surface brightness $I_0$ can be reduced by about $10^8$ times (example of $2.5$m telescope operating at $\lambda=750$nm). Surface brightness of the glare at the apparent distance twice the radius of the disc from the disc center is about $10^{-10}I_0$.

\begin{figure}
  \centering
  \includegraphics[width=175mm]{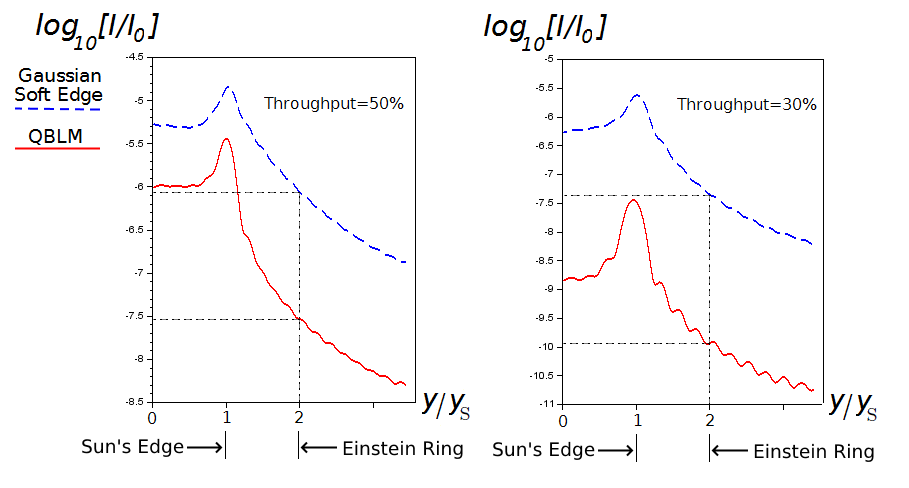}
  \caption{Result of direct numerical simulations for the uniform disc ($\log_{10}$ of relative surface brightness of the sun diffraction glare vs radial coordinate in the coronagraph's detector plane. No limb darkening is taken into account). Dashed/solid curves stand for to the gaussian soft edge/ quasi band limited masks correspondingly. Left Image: throughput $\approx$ 50 percents for both masks. Right image: throughput $\approx$ 30 percents. For details see Appendix 3.}
  \label{fig_qblm_vs_g}
\end{figure}

Our mask shows superior performance in comparison with the Gaussian Soft Edge Mask (GSEM) introduced earlier for the SGL imaging. For instance, for masks with 30 percent throughput, the suppression rate of sunlight by our mask is more than two orders of magnitude better than that by the GSEM (see Figure \ref{fig_qblm_vs_g} for example of 2.5m telescope operating at 2200AU and 750nm wavelength and Appendix 3 for comparative analysis). We note that both GSEM and QBLM are ``amplitude" (i.e. non phase changing on transmission) masks and, therefore, the manufacturing processes for both of them are of the same complexity.

\begin{figure}
  \centering
  \includegraphics[width=125mm]{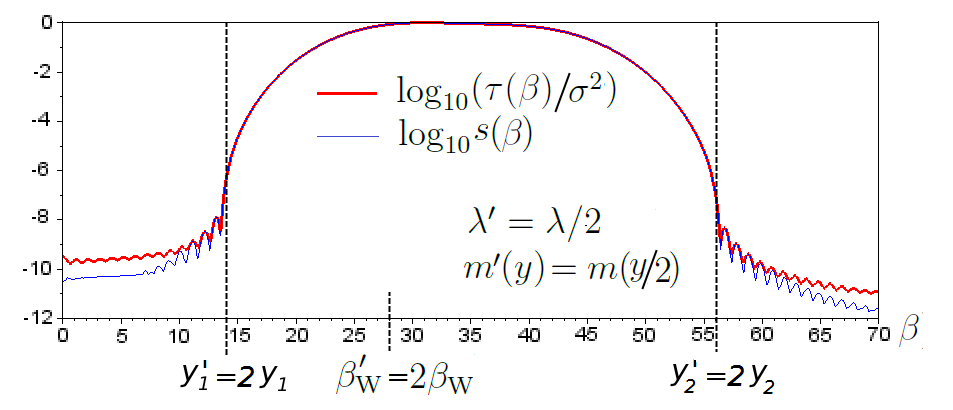}
  \caption{$\log_{10}$ of suppression coefficients of the circularly symmetric mask designed for wavelength $\lambda$ and operating at $\lambda'=\lambda/2$. Thick curve corresponds to the energy suppression coefficient, while thin line corresponds to the coefficient for maximum of PSF.}
  \label{fig_suppression_scaled}
\end{figure}

It is important to note that, similarly to the band-limited masks, the quasi band-limited mask designed for a wavelength $\lambda$ will work for $\lambda'<\lambda$ with the same working throughput and better suppression rates. Figure \ref{fig_suppression_scaled} illustrates the above statement for our example of the circularly symmetric mask designed for $\lambda=750$nm and operating at $\lambda'=\lambda/2=375$nm (compare with Figure \ref{fig_suppression}).

Examples considered in this paper are centered around the particular application: imaging of exo-planets with the help of the solar gravitation lens (SGL). Such imaging requires not only suppressing sunlight, but also reducing glare from the parent star of an exo-planet. The product mask, introduced in the previous section, can perform both of the above tasks.

Another problem related to the SGL imaging is to suppress light from the solar corona: Our mask reduces significantly the glare from almost all the corona, except immediate neighborhood of the Einstein ring. Appendix 4 presents the related results. Distribution of the surface brightness of the detector plane image of corona is shown on Figure \ref{fig_Ieff}.

It is worthy to note that in the context of  SGL imaging suppression of the solar glare up to the level of  brightness of the corona $I\sim 10^{-8}I_0$, rather than to that of the image of the Einstein ring ($I\sim10^{-10}I_0$ in our example) might be sufficient. In such a case parameter $\epsilon$ can be decreased to downgrade the suppression level, increasing the useful throughput in exchange (we recall that the maximal diameter of the Lyot stop is $\sigma=1-\epsilon$) or/and decreasing the telescope aperture.

\begin{figure}
  \centering
  \includegraphics[width=175mm]{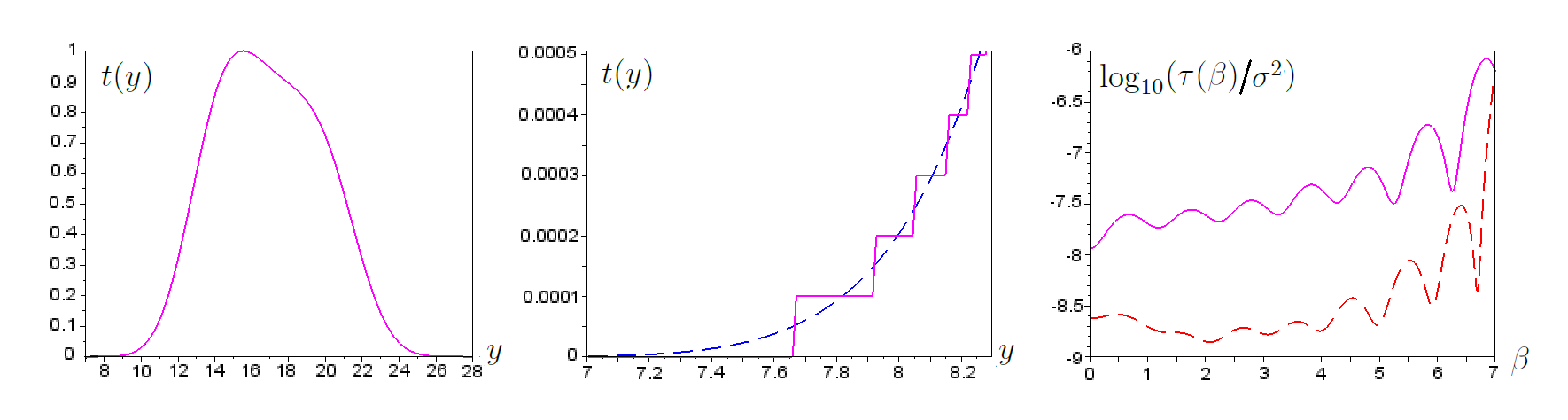}
  \caption{Comparison of the original mask with dicsretized (degraded) mask. Dashed curves show quantities related to the original mask, while solid curves correspond to degraded mask. Left: transparency $t(y)$ across the annular region $y_1<y<y_2$. Center: Zoom of transparency in neighborhood of interior boundary of annulus. Right: Comparison of the degraded energy suppression coefficient with the original one.}
  \label{fig_mask_degrade}
\end{figure}

Concluding the paper we would like to mention the problem of tolerance to manufacturing errors: To test the tolerance of the mask to manufacturing errors we have divided the transparency (also called intensity transmission factor) of the circularly symmetric mask
$$
t(y)=m(y)^2, \quad 0\le t\le 1,
$$
into equidistant discrete gray-scale levels with spacing $\delta t$ between  adjacent levels. Starting from the very small spacing we increased $\delta t$ until the degradation in the level of suppression surpassed about one order of magnitude (which is acceptable for our purposes). This happens when $\delta t$ is of the order of $10^{-4}$. The results of the corresponding simulations are shown on Figure \ref{fig_mask_degrade}. It is worthwhile to mention that our numeric computations were performed on a square spatial grid with spacing $\delta y\approx 0.1$, and, therefore, $\delta t$ is, in fact, the minimal spacing between the levels. The actual spacing at given $y$ is the biggest of the two values $\delta t$ and $|\nabla t(y)\cdot\delta \vec{y}|$. The latter is $\approx 3\times 10^{-3}$ at points where $t(y)$ is steepest.

Another important issue is the question of the wavefront control. Although this question is beyond the scope of the present work, we will make some rough estimates.

According to \cite{KT}, acceptable rms of the wavefront error is approximately proportional to the product of the working incidence and the square root of the suppression ratio. To get an idea whether the wavefront control necessary for the SGL imaging is technically achievable we compare the above mentioned product for our design with that for the WFIRST coronagraph instrument \cite{WFIRST}. Both examples deal with the space telescopes of apertures $\sim 2$ m.

Let us take minimal requirement for our design: suppression to the level of corona. In this case the acceptable energy suppression ratio is between $10^{-6}$ and $10^{-7}$. The working incidence is of order $10^1$. Therefore, the product of the working incidence and the square root of suppression ratio is about of the order of $10^{-2}$.

In the case of the WFIRST instrument, the suppression ratio is between $10^{-8}$ and $10^{-9}$ and the working incidence is between $10^0$ and $10^1$. Therefore the above product is about of the order of $10^{-4}$, which is order(s) of magnitude less than in our case. So, one might expect that the quasi band limited coronagraph can suppress the sun glare to the acceptable level with the currently available wavefront control devices.


Detailed analysis of the wavefront control as well as consideration of QLBM for other (than SGL imaging) applications are the next steps in development of such types of masks: Another application of these masks, one can immediately think of, is their use for observations of faint objects in multiple star systems. A relatively close multiple star system is an extended object that cannot be considered as a single point source and QBLM can simultaneously suppress the light coming from all stars of the system. 


\section{Appendix 1}
\label{pure_bl}

In this Appendix we first derive equation (\ref{Fb}) for the final field of the band-limited coronagraph and then we derive equation (\ref{FQ}) for quasi band limited case.

Since the support of function $A_{\vec{\beta}}(\vec{x})$ is a disc of unit diameter, i.e. $A_{\vec{\beta}}(\vec{x})$ vanishes when $x>1/2$, while $A_{\vec{\beta}}(\vec{x})=E_{\vec{\beta}}(\vec{x})=e^{-2\pi i (\vec{\beta}\cdot\vec{x})}$ for $x<1/2$ (see beginning of Section 2), we have
$$
[M_\epsilon*A_{\vec{\beta}}](\vec{x})=\int_{\rm plane} M_\epsilon(\vec{x}'-\vec{x})A_{\vec{\beta}}(\vec{x}')d^2x'=
\int_{x'<1/2} M_\epsilon(\vec{x}'-\vec{x})A_{\vec{\beta}}(\vec{x}')d^2x'=\int_{x'<1/2} M_\epsilon(\vec{x}'-\vec{x})e^{-2\pi i (\vec{\beta}\cdot\vec{x}')}d^2x'
$$
$$
=\int_{\rm plane} M_\epsilon(\vec{x}'-\vec{x})e^{-2\pi i (\vec{\beta}\cdot\vec{x}')}d^2x'-\int_{x'>1/2} M(\vec{x}'-\vec{x})e^{-2\pi i (\vec{\beta}\cdot\vec{x}')}d^2x'=
$$
$$
=m_\epsilon(\beta)e^{-2\pi i (\vec{\beta}\cdot\vec{x})}-\int_{x'>1/2} M_\epsilon(\vec{x}'-\vec{x})e^{-2\pi i (\vec{\beta}\cdot\vec{x}')}d^2x' .
$$
Therefore ,
\begin{equation}
L(\vec{x})[M_\epsilon*A_{\vec{\beta}}](\vec{x})=m_\epsilon(\beta)L(\vec{x})E_{\vec{\beta}}(\vec{x})-L(\vec{x})\int_{x'>1/2} M_\epsilon(\vec{x}'-\vec{x})e^{-2\pi i (\vec{\beta}\cdot\vec{x}')}d^2x' .
\label{Fb_a}
\end{equation}
The last term in (\ref{Fb_a}) vanishes identically when $\sigma< 1-\epsilon$. In more detail: $L(\vec{x})=0$, when $x>\sigma/2$, so the last term vanishes for $x>\sigma/2$. For $x<\sigma/2$ the factor $M_\epsilon(\vec{x}'-\vec{x})$ in the integrand vanishes. Indeed, the support of $M_\epsilon(x)$ is a disc of diameter $\epsilon$. On the other hand the difference $|\vec{x}'-\vec{x}|>\epsilon/2$, since $x<\sigma/2<(1-\epsilon)/2$, while $x'>1/2$, i.e. the difference $\vec{x}-\vec{x}'$ is outside of the support of $M_\epsilon$ and, therefore , the last term in (\ref{Fb_a}) vanishes identically and we get (\ref{Fb}).

Note that above derivation can be easily generalized for supports of an arbitrary shape (not necessarily discs). The main condition for (\ref{Fb}) to hold is that sum of support of $L$ and that of $M$ should be smaller than support of the aperture function.

Now, we turn our attention to the final field in the quasi-band limited case:
$$
F_{\vec{\beta}}(x)=L(\vec{x})[M*A_{\vec{\beta}}](\vec{x})=L(\vec{x})[M_\epsilon*A_{\vec{\beta}}](\vec{x})+L(\vec{x})[\delta M*A_{\vec{\beta}}](\vec{x})
$$
Taking into account that $L(\vec{x})[M_\epsilon*A_{\vec{\beta}}](\vec{x})=m_\epsilon(\beta)L(\vec{x})E_{\vec{\beta}}(\vec{x})$ and $m_\epsilon=m-\delta m$, we get
$$
F_{\vec{\beta}}(x)=m_\epsilon(\beta)L(\vec{x})E_{\vec{\beta}}(\vec{x})+L(\vec{x})[\delta M*A_{\vec{\beta}}](\vec{x})=m(\beta)L(\vec{x})E_{\vec{\beta}}(\vec{x})+L(\vec{x})\left([\delta M*A_{\vec{\beta}}](\vec{x})-\delta m(\beta)E_{\vec{\beta}}(\vec{x})\right)
$$
which leads to equation (\ref{FQ}).

\section{Appendix 2}

Below we derive (\ref{K_reduced}, \ref{Kernel_1d}): From (\ref{kappa_x}, \ref{kappa_y}) it follows that
$$
\hat{K}_\epsilon[m](\vec{y})=\int_{y_1<y'<y_2}K_\epsilon(\vec{y}-\vec{y'})m(\vec{y'})d^2y'=\int_{y_1<y'<y_2}d^2y'\int_{x<\epsilon/2}d^2xe^{2i\pi \vec{x}\cdot(\vec{y}-\vec{y'})}m(\vec{y'}) .
$$
After integrating in polar coordinates in the $x$-plane only, one can express $K_\epsilon$ in terms of the Bessel function as in eq. (\ref{K_epsilon}), but instead we rewrite all variables in polar coordinates as
$$
\vec{x}=(x\cos\theta,x\sin\theta), \quad \vec{y}=(y\cos\varphi,y\sin\varphi), \quad \vec{y'}=(y'\cos\varphi',y'\sin\varphi') .
$$
Then, taking into account that $m$ is circularly symmetric, i.e. $m(\vec{y'})=m(y')$, and integrating first in $\phi'$ and then in $\theta$, we get
$$
\hat{K}_\epsilon[m](y)=\int_{y_1}^{y_2}\left[(2\pi)^2\int_0^{\epsilon/2}J_0(2\pi xy)J_0(2\pi xy')xdx\right]m(y')y'dy' .
$$
Taking integral in the square brackets, we obtain (\ref{K_reduced}, \ref{Kernel_1d}).

\section{Appendix 3}
\label{compare}


Result of direct numerical simulations for the uniform disc (no limb darkening is taken into account) are shown on Figure \ref{fig_qblm_vs_g}. Here we compare performance of the radially symmetric QLBM with gaussian soft edge mask (GSEM). For the latter
$$
m(y)=\left\{
\begin{array}{l}
0, \quad y\le y_{\rm g}\\
1-\exp[(y-y_{\rm g})^2/g^2], \quad y\ge y_{\rm g}
\end{array}
\right.
$$
We set $y_{\rm g}=7$, which corresponds to the edge of the sun. We compared performance of two masks provided both have equal throughput and resolution (Lyot stop diameter). Throughput of the GSEM at given diameter of the Lyot stop $\sigma=1-\epsilon$ depends on parameter $g$.

In the case of the 30 percent throughput we took example of QLBM considered in the main text (i.e. $y_1=7$, $y_2=28$, $\epsilon=0.4$). Here $g=4.4$ and the suppression rate of QLBM is about 400 times better than that of the GSEM (see right panel of Figure \ref{fig_qblm_vs_g}).

Another example corresponds to the 50 percent throughput:  We took QLBM with $y_1=7.35$, $y_2=25.75$ and $\epsilon=0.25$. Here, $g=4.1$ and the suppression rate of QLBM is about 40 times better than that of the GSEM (see left panel of Figure \ref{fig_qblm_vs_g}).

Note that one may further optimize the GSEM for a given throughput by varying all parameters $y_{\rm g}$, $\sigma$ and $g$ under condition that $\left(1-\exp[(\beta_{\rm W}-y_{\rm g})^2/g^2]\right)\sigma$ is fixed. For instance, in the case of the 30-percent throughput one could try $\sigma=.75$ and $g=6.1$ (instead of $\sigma=.6$ and $g=4.4$). However numerical result show that such optimization does not lead to significant improvements in performance of the GSEM.

Suppression of the solar glare up to the level of  brightness of corona, rather than to that of the image of the Einstein ring (as in the main text example) might be sufficient. From the results presented one can see that QLBM can perform such a task \footnote{We recall that taking limb darkening into account leads to about double reduction of the glare brightness} at about 50 percent throughput, while throughput of the GSEM will be about twice smaller and the corresponding image resolution will be also rougher.

\section{Appendix 4}

With great precision eq. (\ref{psf}) holds for incidences corresponding to the light from corona $\beta>y_1$. Taking into account (\ref{brightness_ps}) and (\ref{brightness}) we obtain surface brightness of the corona's glare in the detector plane
\begin{equation}
I(\vec{y})=\int I_{\rm c}(\beta)m^2(\beta)\mathcal{I}_{\rm L}(\vec{\beta},\vec{y})d^2\beta ,
\label{brightness_corona}
\end{equation}
where $I_c(\beta)$ is the distribution of the surface brightness in corona and $\mathcal{I}_{\rm L}(\vec{\beta},\vec{y})=\mu_{\rm Lyot}(\vec{\beta},\vec{y})/S $.
\begin{figure}
  \centering
  \includegraphics[width=115mm]{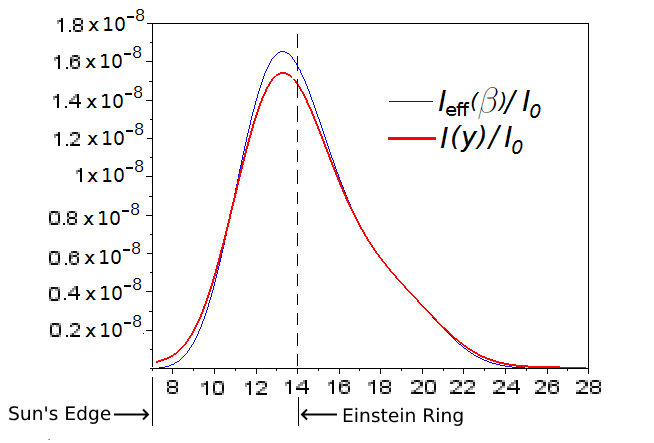}
  \caption{Thick curve: Dependence of distribution of the relative surface brightness of the detector plane image of corona on $y$, i.e. $I(y)/I_0$. Thin curve: Dependence of $I_{\rm eff}(\beta)/I_0$ on $\beta$.}
  \label{fig_Ieff}
\end{figure}
From (\ref{brightness_corona}) it follows that effect of mask is equivalent to reducing distribution of the surface brightness of corona by the factor $m(\beta)^2$. In other words, corona is seen in the detector plane as if there were no mask, but instead the surface brightness distribution of the corona were $I_{\rm eff}(\beta)=m(\beta)^2 I_{\rm c}(\beta)$.

According to \cite{NK}, dependence of the distribution $I_c$ on $\beta$ is
\begin{equation}
I_c=\left(\frac{3.67}{\rho^{18}}+\frac{1.939}{\rho^{7.8}}+\frac{0.0551}{\rho^{2.5}}\right)\times 10^{-6}I_0,
\label{Ic}
\end{equation}
where
$$
\rho=\beta/\beta_{\rm S}>1
$$
is the distance from the center of the sun in units of angular radius of the sun. In our example $\beta_{\rm S}=y_1$. Multiplying (\ref{Ic}) and $m(\beta)^2$ from our example of the circularly symmetric mask (corresponding $m(\beta)$ is shown on Figure \ref{fig_m}), we get effective distribution of brightness $I_{\rm eff}$ shown on Figure \ref{fig_Ieff}. Figure \ref{fig_Ieff} also shows distribution of the surface brightness of image of corona in the detector plane. Action of the mask results in significant effective reduction of the corona brightness almost everywhere, except close neighbourhood of the Einstein ring. (compare Figure \ref{fig_Ieff} with Figure \ref{fig_SGL})


\end{document}